\documentclass[10pt,conference]{IEEEtran}
\usepackage{cite}
\usepackage{amsmath,amssymb,amsfonts}
\usepackage{graphicx}
\usepackage{textcomp}
\usepackage{xcolor}
\usepackage[hyphens]{url}
\usepackage{fancyhdr}
\usepackage{hyperref}

\newcommand{\hpcayear}{2026}

\usepackage{algorithm}
\usepackage[noend]{algpseudocode}

\newcommand{\T}{\textit{FractalCloud}}
\usepackage{xcolor}
\usepackage{multirow}
\usepackage{booktabs}
\usepackage{multicol}
\usepackage{makecell}
\usepackage{graphicx}
\graphicspath{{micro58-latex-template/figs/}}
\usepackage{textcomp}
\usepackage{xcolor}
\usepackage{subfig}
\usepackage[labelformat=empty]{subcaption}
\usepackage[hyphens]{url}

\usepackage[normalem]{ulem} 
\newcommand{\retagNoLink}[1]{\hyperref[#1]{\textcolor{brown}{\textbf{#1}}}} 
\newcommand{\retag}[1]{\hyperref[#1]{\textcolor{brown}{\underline{\textbf{#1}}}}}
\newcommand{\revise}[1]{{\color{blue}#1}}
\newcommand{\revision}[2]{\phantomsection\label{#1}\textbf{\emph{\hyperref[CoverLetter]{\textcolor{brown}{\underline{#1}}}}} \textcolor{blue}{#2}} 
\newcommand{\recaption}[2]{\textbf{\emph{\hyperref[CoverLetter]{\textcolor{brown}{\underline{#1}}}}} \textcolor{blue}{#2}}

\def\hidebrown{}

\ifdefined\hidebrown
  \renewcommand{\retagNoLink}[1]{}
  \renewcommand{\retag}[1]{}
  \renewcommand{\revision}[2]{#2}
  \renewcommand{\recaption}[2]{#2}
  \renewcommand{\revise}[1]{{#1}}
\fi

\newcommand{\hpcasubmissionnumber}{1780}
\title{FractalCloud: A Fractal-Inspired Architecture for Efficient Large-Scale Point Cloud Processing} 

\def\hpcacameraready{} 

\newcommand\hpcaauthors{Yuzhe Fu$^\dagger$, Changchun Zhou$^\dagger\textsuperscript{*}$, Hancheng Ye$^\dagger$, Bowen Duan$^\dagger$, Qiyu Huang$^\ddagger$, Chiyue Wei$^\dagger$, Cong Guo$^\dagger\textsuperscript{*}$, \\Hai ``Helen'' Li$^\dagger$, and Yiran Chen$^\dagger$}
\newcommand\hpcaaffiliation{$^\dagger$Duke University, $^\ddagger$Yale University}
\newcommand\hpcaemail{\{yuzhe.fu, changchun.zhou, cong.guo, hai.li, yiran.chen\}@duke.edu}

\author{
  \ifdefined\hpcacameraready
    \IEEEauthorblockN{\hpcaauthors{}}
      \IEEEauthorblockA{
        \hpcaaffiliation{} \\
        \hpcaemail{}
      }
  \else
    \IEEEauthorblockN{\normalsize{HPCA \hpcayear{} Submission
      \textbf{\#\hpcasubmissionnumber{}}} \\
      \IEEEauthorblockA{
        Confidential Draft \\
        Do NOT Distribute!!
      }
    }
  \fi 
}

\fancypagestyle{camerareadyfirstpage}{%
  \fancyhead{}
  
  \fancyhead[C]{
    \ifdefined\aeopen
    \parbox[][12mm][t]{13.5cm}{\hpcayear{} IEEE International Symposium on High-Performance Computer Architecture (HPCA)}    
    \else
      \ifdefined\aereviewed
      \parbox[][12mm][t]{13.5cm}{\hpcayear{} IEEE International Symposium on High-Performance Computer Architecture (HPCA)}
      \else
      \ifdefined\aereproduced
      \parbox[][12mm][t]{13.5cm}{\hpcayear{} IEEE International Symposium on High-Performance Computer Architecture (HPCA)}
      \else
      \parbox[][0mm][t]{13.5cm}{\hpcayear{} IEEE International Symposium on High-Performance Computer Architecture (HPCA)}
    \fi 
    \fi 
    \fi 
    \ifdefined\aeopen 
      \includegraphics[width=12mm,height=12mm]{ae-badges/open-research-objects.pdf}
    \fi 
    \ifdefined\aereviewed
      \includegraphics[width=12mm,height=12mm]{ae-badges/research-objects-reviewed.pdf}
    \fi 
    \ifdefined\aereproduced
      \includegraphics[width=12mm,height=12mm]{ae-badges/results-reproduced.pdf}
    \fi
  }
  \fancyfoot[C]{}
}
\begin{document}


\maketitle
\renewcommand{\thefootnote}{\fnsymbol{footnote}}
\footnotetext[1]{Corresponding Author: Changchun Zhou and Cong Guo}
\renewcommand{\thefootnote}{\arabic{footnote}}

\ifdefined\hpcacameraready 
  \thispagestyle{camerareadyfirstpage}
  \pagestyle{empty}
\else
  \thispagestyle{plain}
  \pagestyle{plain}
\fi

\newcommand{\hpcaheight}{0mm}
\ifdefined\eaopen
\renewcommand{\hpcaheight}{12mm}
\fi


\begin{abstract}

Three-dimensional (3D) point clouds are increasingly used in applications such as autonomous driving, robotics, and virtual reality (VR). Point-based neural networks (PNNs) have demonstrated strong performance in point cloud analysis, originally targeting small-scale inputs. However, as PNNs evolve to process large-scale point clouds with hundreds of thousands of points, \revision{RB1a}{all-to-all computation} and global memory access in point cloud processing introduce substantial overhead, causing \revision{RB1b}{$O(n^2)$ computational complexity and memory traffic where $n$ is the number of points}. Existing accelerators, primarily optimized for small-scale workloads, overlook this challenge and scale poorly due to inefficient partitioning and non-parallel architectures. To address these issues, we propose \textit{\T}, a fractal-inspired hardware architecture for efficient large-scale 3D point cloud processing. \textit{\T} introduces two key optimizations: \revision{RC1b}{(1) a co-designed Fractal method for shape-aware and hardware-friendly partitioning}, and (2) block-parallel point operations that decompose and parallelize all point operations. A dedicated hardware design with on-chip fractal and flexible parallelism further enables fully parallel processing within limited memory resources. Implemented in 28 nm technology as a chip layout with a core area of 1.5 $mm^2$, \textit{\T} achieves 21.7× speedup and 27× energy reduction over state-of-the-art accelerators while maintaining network accuracy, demonstrating its scalability and efficiency for PNN inference. The code for \textit{\T} is available at \url{https://github.com/Yuzhe-Fu/FractalCloud}.

\end{abstract}

\section{Introduction}

With the rapid advancement of 3D sensing technologies~\cite{rani2024advancements, guo2024revisiting, yan2024spot, zhou2023sagitta}, such as LiDAR~\cite{roriz2024survey} and stereo cameras~\cite{dong20231920, dong20214, dong2021139}, point cloud has become prevalent in a wide range of edge applications, including autonomous driving~\cite{liu2025towards, cui2021deep}, robotics~\cite{kastner20203d}, augmented reality~\cite{wang2018point}, and virtual reality~\cite{stets2017visualization, li2024fusion}.
The points in a 3D point cloud are inherently unordered and irregularly distributed, making it challenging for conventional deep neural networks (DNNs) to directly extract meaningful local features~\cite{ali2018yolo3d, yang2018pixor, zhang2023pids, choy20194d}.
To overcome this limitation, point-based neural networks (PNNs)~\cite{he2025pointrwkv, deng2023pointvector, qian2022pointnext, qi2017pointnet++, qi2017pointnet,  phan2018dgcnn, zhao2019pointweb, wang2019dynamic} incorporate specialized point operations, such as sampling, neighbor searching, and gathering, to effectively extract learnable local features from unordered point cloud. These tailored operations significantly improve the performance of PNNs in 3D perception tasks.
However, these operations often involve irregular memory access patterns and intensive computation, which pose significant challenges to the efficiency of PNNs on edge hardware.

Numerous architecture researchers have proposed various hardware- and software-level optimizations to accelerate PNN execution, aiming to improve both throughput and energy efficiency.
Early works, such as Mesorasi~\cite{feng2020mesorasi} and PointAcc~\cite{lin2021pointacc}, primarily focused on optimizing intensive tensor computations (i.e., DNN layers), which were the main performance bottleneck at the time.
As these tensor-level bottlenecks were mitigated, the bottleneck shifted to the irregular memory access from point operations.
Consequently, more efforts have focused on optimizing memory access efficiency. One common approach is to partition data into blocks and organize them to alleviate memory access conflicts~\cite{feng2022crescent}. 
\revision{RC1a}{Earlier algorithmic work~\cite{talbert2000empirical, moore1990efficient} analyzed various partitioning methods (e.g., KD-trees) on conventional datasets without considering hardware optimization or the geometric structures of point-cloud data. 
In contrast, recent architectural designs} such as PNNPU~\cite{kim2021pnnpu} employ a space-uniform partitioning strategy, dividing the 3D space by point coordinates to generate structured memory access patterns. Crescent~\cite{feng2022crescent} adopts a density-uniform partitioning scheme based on the KD-tree to achieve more balanced workloads and improved memory locality. 

\begin{figure}[t]
    \centering
    \includegraphics[width=0.9\linewidth]{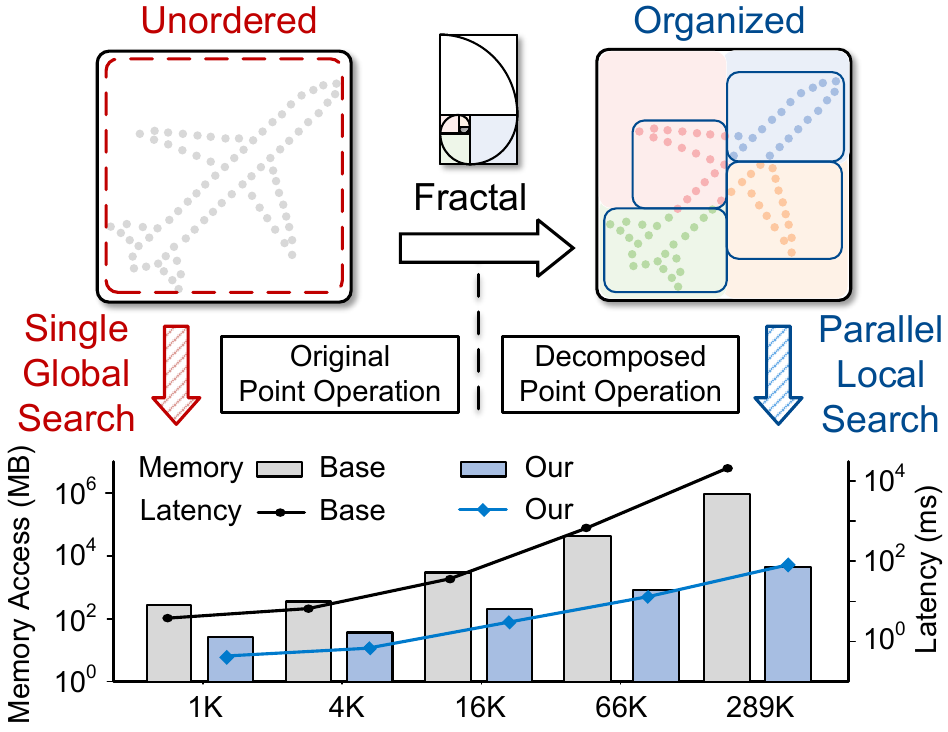}
    \caption{The memory access (MB) and inferencing latency (ms) of point cloud neural networks in (a) original baseline structure and (b) the proposed \textit{\T}.}
    \label{fig:intro}
\end{figure}

However, as PNN evolves, the scale of point cloud data has expanded from 1K to over 200K in benchmarks~\cite{sun2022modelnet40,armeni20163d}, with modern LiDAR sensors producing 30K–300K points per frame in real life~\cite{roriz2024survey, li2023instant}. This increase introduces new challenges, with point operations emerging as the primary bottleneck, as shown in Fig.~\ref{fig:GPULatency}. \revision{RB2b}{Conventional point operations require iterative global memory access and \revision{RB1b}{all-to-all computation}, referred to as global search. The iterative nature and data dependency severely limit decomposition and parallelization}, both essential for efficient large-scale processing~\cite{guo2020accelerating, guo2024accelerating}. Partitioning mitigates global search by localizing computations and memory accesses. However, prior works mainly target small-scale inputs ($<4K$ points) and overlook this parallelization issue. As a result, their hardware designs are misaligned with the demands of modern workloads, causing significant performance degradation on large-scale point clouds.

After detailed analysis, we identify two fundamental challenges in state-of-the-art (SOTA) accelerators:

\begin{figure}[t]
    \centering
    \includegraphics[width=0.9\linewidth]{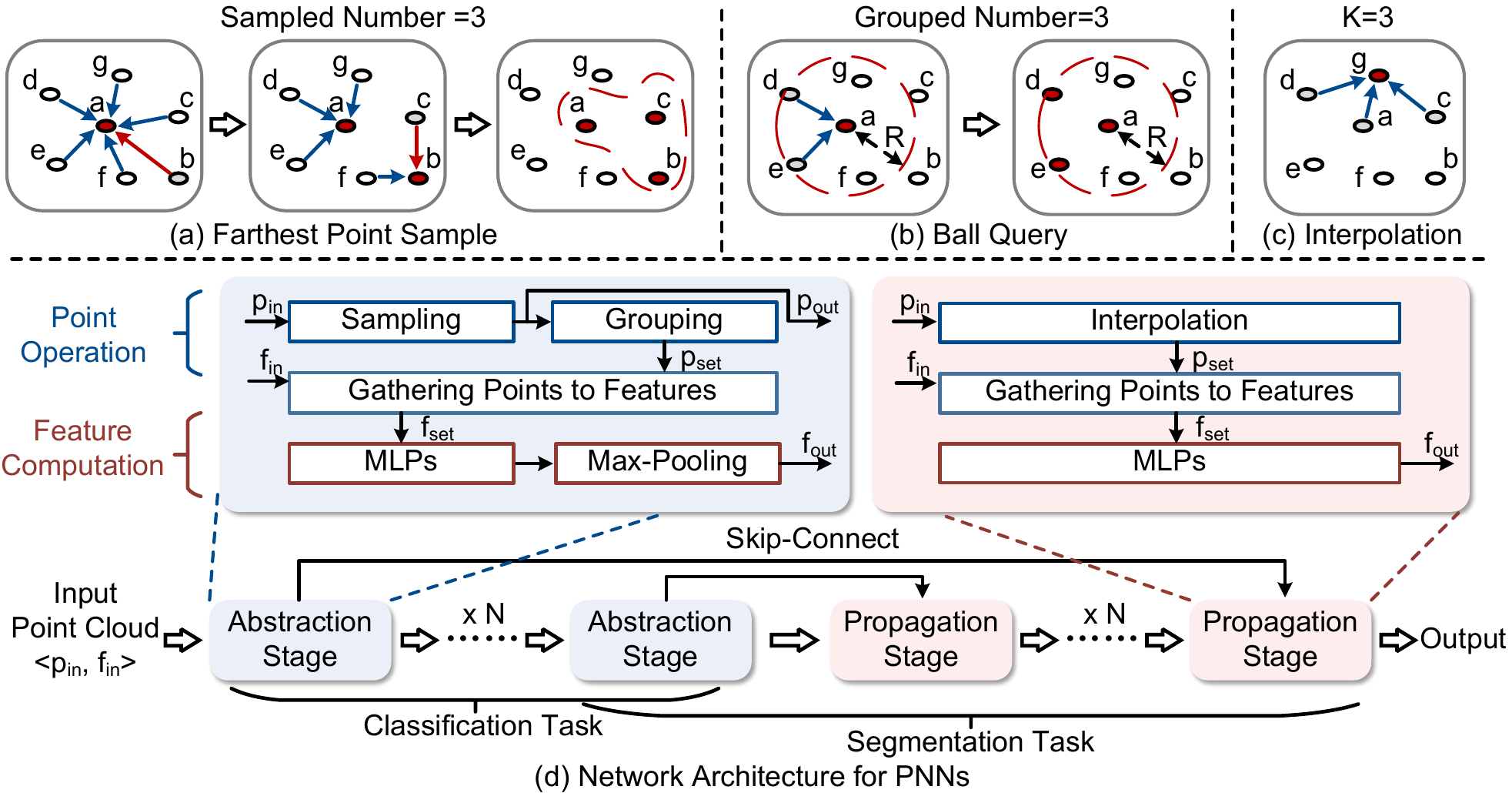}
    \caption{The workflows for (a) farthest point sampling, (b) ball query, and (c) interpolation. (d) The backbone of PNNs.}
    \label{fig:network}
\end{figure}

\textbf{Inefficient Partitioning.} Existing partitioning strategies exhibit a fundamental trade-off between hardware efficiency and network accuracy. Simple space-uniform partitions~\cite{kim2021pnnpu} are hardware-friendly but ignore highly uneven point distributions, leading to imbalanced blocks and severe accuracy degradation~\cite{zhou2024adjustable}, making them unsuitable for large-scale PNNs. Density-aware partitions, such as the KD-tree-based method~\cite{feng2022crescent,pinkham2020quicknn}, achieve balanced partitions that preserve accuracy but rely on recursive and exclusive sorting operations. These sorts are inherently sequential and non-decomposable, creating severe hardware inefficiency and becoming a new bottleneck (e.g., 53\% latency in Crescent), making them inefficient with increased point cloud.

\textbf{Non-Parallel Architecture.} Existing architectures lack optimizations on block-level parallelism. \revision{RB2a}{Accelerators without partitioning~\cite{feng2020mesorasi, lin2021pointacc} would make non-decomposable point operations with long latency}, as shown in Fig.~\ref{fig:intro}. Even accelerators with partitioning remain constrained by the sequential nature of partitioning methods~\cite{feng2022crescent}, resulting in block-serial execution. While acceptable for small-scale inputs (e.g., Crescent is only 20\% slower than ours), this design assumption fails for large-scale point clouds. As input size \revision{RB1b}{$n$} increases, partitioning latency grows rapidly, and the lack of block-level parallelism causes latency to scale almost \revision{RB1b}{$O(n^2)$}, severely limiting overall performance.

\textbf{Main Idea.} To enable efficient large-scale point cloud processing and address the two fundamental challenges, we propose \textit{\T}, a fractal-inspired hardware architecture. It organizes point cloud into blocks with a fractal strategy and localizes global point operations into fully parallel block-wise execution, as shown in Fig.~\ref{fig:intro}. Our contributions are summarized as follows:

\begin{itemize}
\item \textbf{Fractal}: \revision{RC1b}{A hardware-driven, shape-aware partitioning method with a DFT-based memory layout and threshold-controlled block division, enabling efficient large-scale point cloud processing.}

\item \textbf{Block-Parallel Point Operations}: Decomposing all point operations from global to local search, enabling fully parallel block-wise processing, and significantly reducing memory accesses and latency.

\item \textbf{Dedicated Hardware Architecture}: A domain-specific accelerator optimized for fractal-based processing, featuring on-chip fractal, flexible parallelism, and data reuse to accelerate parallelized point operations.

\item In 28 nm technology, our \textit{\T} achieves $1.5 mm^2$ and $0.58W$, delivering an average 21.7× speedup and 27× energy reduction over state-of-the-art accelerators.

\end{itemize}

\section{Background}
In this section, we firstly show the backbone of typical PNNs, then explain the point operations in PNNs.

\subsection{The Backbone of PNNs}

A point cloud consists of two types of information: spatial coordinates $p$ and feature information $f$. As illustrated in Fig.~\ref{fig:network}(d), a typical PNN architecture comprises two main stages: Abstraction and Propagation. Each stage includes two sequential computational pathways: Point Operations, performed on spatial coordinates to extract learnable local geometric features, and Feature Computation, which processes and aggregates these features to learn high-level semantic representations~\cite{deng2023pointvector, qian2022pointnext, qi2017pointnet++, qi2017pointnet,  phan2018dgcnn, zhao2019pointweb, wang2019dynamic}.

In classification tasks, PNNs consist of the Abstraction stages. Firstly, the spatial coordinates $p_{in}$ of the input point cloud undergo point operations, where a sampling operation selects representative points. A grouping operation then identifies spatially neighboring points for each sampled point, forming local point sets $p_{set}$, which are subsequently transformed into the feature space through a gathering operation, producing Feature Sets $f_{set}$. $f_{set}$ are then processed in the Feature Computation, which applies Multi-Layer Perceptrons (MLPs) to extract hierarchical features. 

For more complex segmentation tasks, PNNs introduce additional Propagation stages that are skip-connected to previous Abstraction stages to recover features. The interpolation operations are used to restore the features of points discarded in the previous sampling steps. It leverages the features of sampled points and reconstructs the pre-sampling point cloud information. This process effectively recovers detailed spatial structures and enhances segmentation accuracy.

\begin{figure*}[t]
    \centering
    \includegraphics[width=0.9\linewidth]{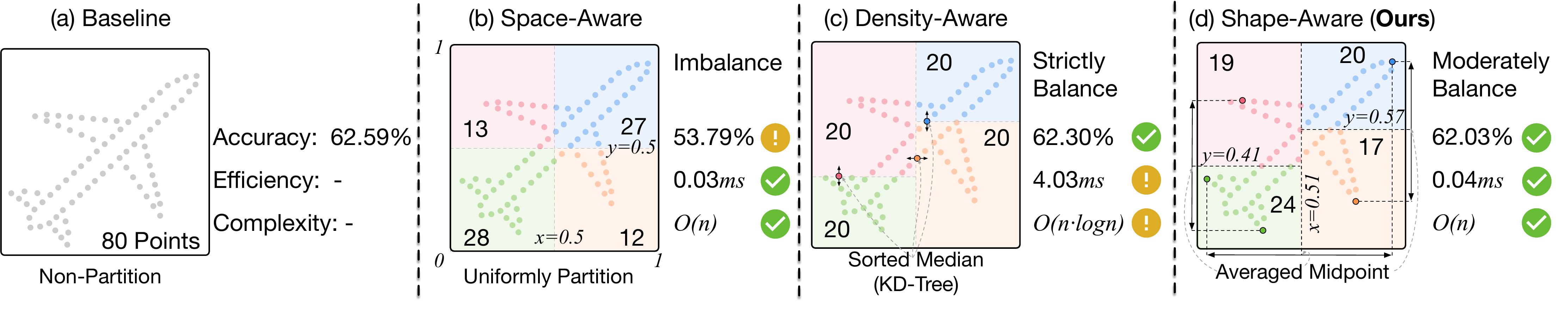}
    \caption{Comparison of different partitioning strategies on latency and network accuracy for PointNeXt segmenting S3DIS dataset across (a) the original point cloud (PointAcc~\cite{lin2021pointacc}), (b) uniform partitioning (PNNPU~\cite{kim2021pnnpu}), (c) KD-tree-based partitioning (Crescent~\cite{feng2022crescent}), and \recaption{RA3}{(d) the proposed Fractal method}.}
    \label{fig:DiffPartition}
\end{figure*}

\subsection{Point Operations in PNNs}

\textbf{Sampling.} \revision{RB5}{Farthest Point Sampling (FPS) is the dominant sampling method in both SOTA PNNs~\cite{qian2022pointnext, deng2023pointvector, he2025pointrwkv} and hardware accelerators~\cite{lin2021pointacc, yoon2023efficient, gao2024hgpcn} for its strong representativeness and accuracy for point cloud processing}, as shown in Fig.~\ref{fig:network} (a).
The process starts with a randomly selected initial point as the first sampled point in the sampled set. In each iteration, the algorithm computes the Euclidean distance between each unsampled point and the sampled set, selecting the farthest unsampled point for inclusion of the sampled set. 

\textbf{Neighbor Searching.} Neighbor searching selects local points around designated center points for grouping and interpolation. PNNs commonly adopt the Ball Query (BQ) method for grouping, which selects up to $num$ points within a radius $R$ based on Euclidean distance as shown in Fig.~\ref{fig:network}(b). For interpolation, the K-Nearest Neighbors (KNN) method is used, retrieving the $K$ closest points without radius constraints by searching the entire candidate set, as shown in Fig.~\ref{fig:network}(c).

\textbf{Gathering.} The gathering operation follows the neighbor searching operations, such as grouping and interpolation. It leverages the neighbor indices obtained from these operations to retrieve corresponding feature values from the point cloud's feature space. These gathered features are then passed to the subsequent MLP module for further feature computation.

\revision{RB4}{\textbf{Complexity.} 
As shown in Fig.~\ref{fig:network}(a–c), original FPS and neighbor searching are iterative with $O(n)$ iterations, each performing an all-to-all computation over $n$ points, resulting in a $O(n^2)$ computation complexity and memory traffic. Gathering also requires frequent global data accesses to retrieve scattered features due to the random layout in original point cloud.}

\section{Motivation}

\subsection{Bottleneck Shift: From Tensor to Point Operation}
\begin{figure}[tb]
    \centering
    \includegraphics[width=1\linewidth]{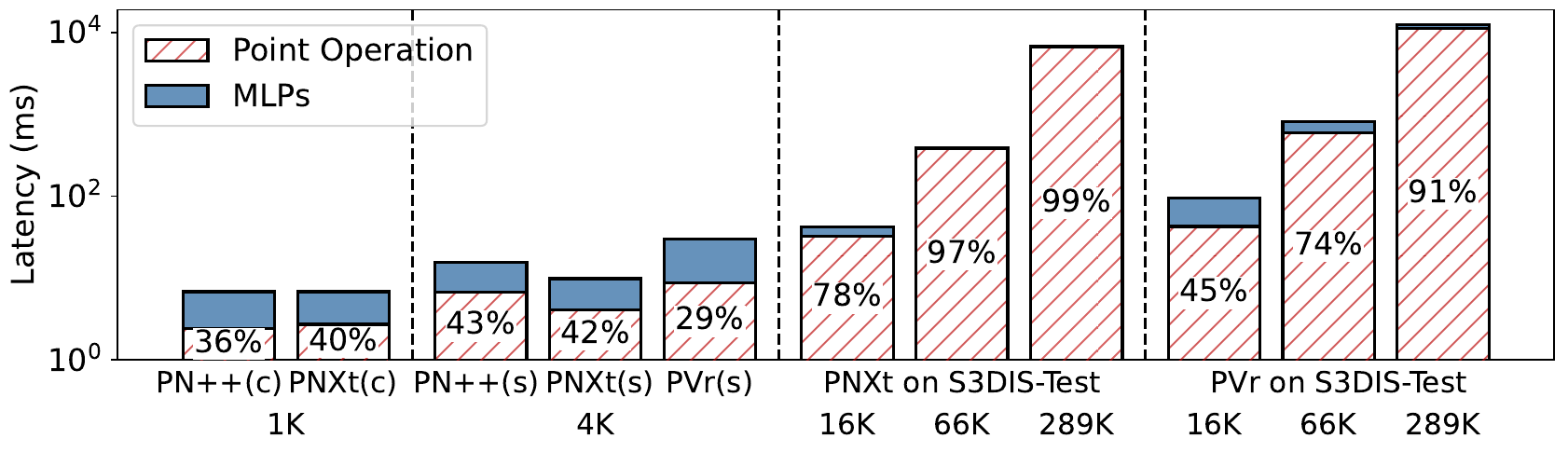}
    \caption{The latency for PN++, PNXt, and PVr \cite{qi2017pointnet++, qian2022pointnext, deng2023pointvector} inference on GPU for classification and segmentation workloads, with notations defined in Table~\ref{tab:evaluatedModels}, under varying input points.}
    \label{fig:GPULatency}
\end{figure}

In recent years, the latency of point-based neural networks (PNNs) has escalated significantly with increasing point cloud sizes and network complexity. 
As shown in Fig.~\ref{fig:GPULatency}, the primary performance bottleneck has shifted from tensor computations to point operations. 
Unlike tensor computations, which can be efficiently parallelized into small, localized workloads, point operations often require global search over the entire point cloud, making them inherently difficult to parallelize. 
For example, sampling techniques such as FPS involve sequential and iterative global traversals to select representative points, while neighbor searching scans the full point set to identify local structures from unordered data. 
These global operations incur $O(n^2)$ computational complexity and memory traffic, causing point operations to dominate overall latency—rising from 30\% at 1K points to over 90\% at 289K points.

This bottleneck shift is also evident in SOTA accelerators~\cite{feng2020mesorasi, lin2021pointacc, feng2022crescent}, which primarily target small-scale point clouds (less than 4K points), where the cost of point operations is less pronounced. 
As a result, these designs focus on optimizing MLP computation and memory streaming, while leaving the global-search inefficiencies of point operations unaddressed.
When applied to large-scale point clouds, such as the S3DIS dataset~\cite{armeni20163d}, existing accelerators suffer from limited on-chip storage, \revision{RA1a}{causing 51\% average cache misses} and frequent DRAM accesses for global search, which further exacerbate latency and expose inefficiencies in point operations.

\textbf{Takeaway}. Point operations have emerged as the dominant performance bottleneck (accounting for $>90\%$ of runtime) in large-scale PNNs, underscoring the urgent need for architectural support and algorithmic innovations targeting their global-search nature.

\subsection{Algorithm: Accuracy and Efficiency Trade-off} 
A variety of PNN acceleration studies have proposed partitioning strategies capable of mitigating the global-search bottleneck, as illustrated in Fig.~\ref{fig:DiffPartition}. 
These approaches include the baseline design~\cite{lin2021pointacc}, space-aware partitioning~\cite{kim2021pnnpu}, density-aware partitioning~\cite{feng2022crescent}, and our proposed method—\textbf{Fractal}, a shape-aware partitioning scheme.

\textbf{Baseline.}
As discussed in the previous subsection, early-stage PNN architectures retain full accuracy~\cite{feng2020mesorasi,lin2021pointacc}, but suffer from poor efficiency due to irregular and memory-intensive point operations. 
To address this limitation, researchers have introduced partitioning schemes aimed at enhancing data locality and streaming memory access.

\textbf{Space-Uniform Partition.}
As shown in Fig.~\ref{fig:DiffPartition}(b), space-aware partitioning uniformly divides the 3D space by spatial coordinates~\cite{kim2021pnnpu,zhou2024adjustable}.
It introduces negligible overhead ~\revision{RA1b}{with a single global traversal}, but ignores the non-uniform distribution of real-world point clouds. This leads to highly imbalanced point densities across blocks, weakening local context and causing notable accuracy loss in subsequent point operations ($>9\%$ in PointNeXt~\cite{zhou2024adjustable}). 
Thus, space-uniform partitioning is inadequate for high-accuracy applications. 

\textbf{Density-Uniform Partition.}
To address the workload imbalance and preserve model accuracy, density-aware partitioning (DAP) methods, such as KD-tree-based partition used in Crescent~\cite{feng2022crescent}, generate balanced partitions by recursively splitting the point cloud based on density. 
As depicted in Fig.~\ref{fig:DiffPartition}(c), KD-tree construction requires recursive median sorting, resulting in high computational complexity~\cite{xu2019tigris} \revision{RA1c}{and accounting for 53\% of Crescent’s latency~\cite{feng2022crescent}.}
Although KD-trees work reasonably well for small-scale point clouds optimization, their performance deteriorates significantly as the input scale increases by 2--3 orders of magnitude. For instance, with 1K points, KD-tree based method is only 20\% slower than our proposed \textit{\T}. 
However, for large-scale inputs (e.g., 289K points), it becomes nearly 100$\times$ slower than \textit{\T} due to poor algorithm-hardware alignment, as further discussed in the next subsection.

\textbf{Shape-Uniform Partition.}
To overcome the limitations of prior methods, we propose the \textbf{Shape-Aware Partition} method, inspired by fractal geometry, and refer to it as \textbf{Fractal}. 
As shown in Fig.~\ref{fig:DiffPartition}(d), \textit{\T} performs linear-time traversal to identify marginal points along each dimension and computes axis-wise midpoints for recursive partitioning. 
This approach leverages object shape information to generate partitions with minimal computational overhead \revision{RA1d}{($<0.8\%$ latency in \textit{\T})} compared to KD-tree methods. 
Moreover, \textit{\T} adaptively partitions dense regions into finer-grained blocks to avoid extreme imbalance, achieving moderately balanced partitions with negligible accuracy loss ($<0.6\%$) and significantly improved processing efficiency.

\textbf{Takeaway.} The core insight behind Fractal is that point distributions often align with the object’s geometric shape due to consistent sampling frequency. 
By being shape-aware, \textit{\T} achieves an effective balance between efficiency and accuracy, making it well-suited for scalable, high-performance PNN processing.

\begin{figure}[tb]
    \centering
    \includegraphics[width=0.9\linewidth]{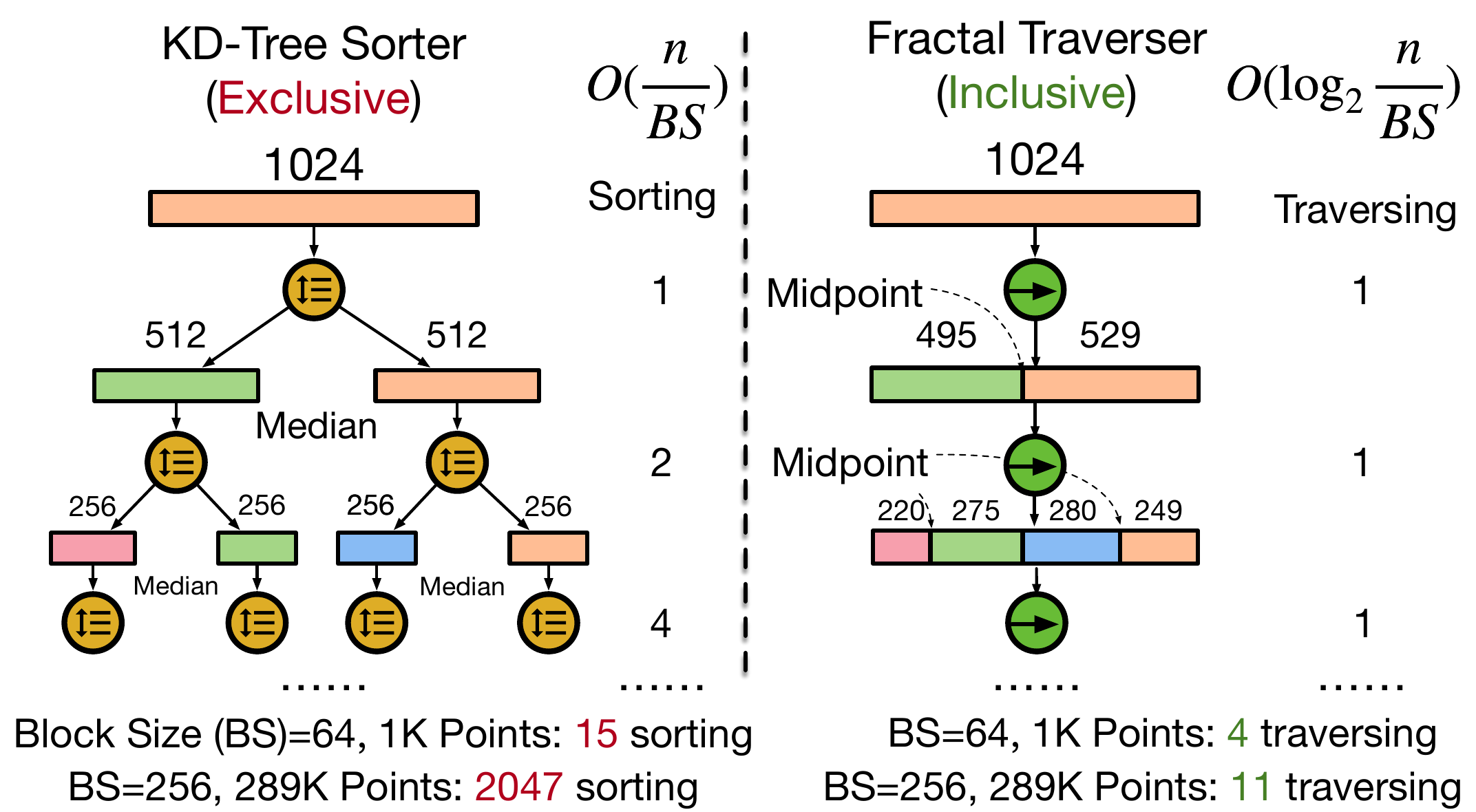}
    \caption{\recaption{RA9}{Workflows for KD-tree and proposed Fractal}. BS is the block size, indicating the max points per block.}
    \label{fig:hardware_efficiency}
\end{figure}

\subsection{Architecture: Being Friendly to Hardware} 
The Fractal method significantly outperforms density-aware partitionings (DAP) in hardware efficiency, primarily due to its structural compatibility with parallel hardware architectures.

\textbf{Exclusive Sorter.} Density-aware partitioning (DAP) methods, such as KD-tree, rely on recursive median sorting as shown in Fig.~\ref{fig:hardware_efficiency} left. 
Each sort in the KD-tree acts as an exclusive, indivisible functional operation, making it inherently sequential and difficult to parallelize or pipeline efficiently. 
For example, when processing 1024 points, a 1024-point sorter yields a global order that cannot be directly decomposed or reused to execute two subsequent local 512-point sorts in parallel.
And all sorts in KD-tree are mandatory as KD-tree’s changing subsets and axes invalidate prior results, requiring re-sorting at each iteration.
Moreover, deeper iterations progressively increase sorts with smaller workloads, which mismatches hardware optimized for large sorting and leads to low resource utilization. 
Consequently, DAP incurs substantial computational overhead, with $O(n/BS)$ complexity, where $BS$ is the final block size, and becomes a primary bottleneck (e.g., 53\% latency in Crescent~\cite{feng2022crescent}).

However, for small-scale inputs, this bottleneck has limited impact.
Even with block-serial processing, existing accelerators achieve acceptable performance (only 20\% slower than ours). 
As a result, prior work largely overlooked block-level parallelism. 
In large-scale scenarios, this assumption breaks down: KD-tree requires 2047 serial sorts for 289K points, and the lack of block-level parallelism further amplifies overall latency.
As a result, DAP severely limits the parallelism of point cloud processing, posing a critical bottleneck for high-throughput accelerators where scalable parallel execution is essential.

\textbf{Inclusive Traverser.}
In contrast, the proposed Fractal adopts a lightweight, linear-time traversal to identify marginal points and determine partitioning midpoints. As shown in Fig.~\ref{fig:hardware_efficiency} right, only four traversals are required to partition 1024 points, and even for 289K points the process completes in just 11 traversals, a sharp contrast to the thousands of recursive sorts in KD-tree. The traversal unit is inherently simple and highly parallelizable. Analogous to arithmetic components like adders, it can be decomposed into concurrent sub-tasks operating across point dimensions or blocks. This design fundamentally aligns with the requirements of parallel-processing architectures, reducing complexity to $O(\log_2 n / BS)$ and enabling scalable, low-latency, and hardware-friendly partitioning for large-scale point clouds. Moreover, by producing balanced and independently processable blocks, Fractal provides the essential foundation for block-level parallelism in subsequent point operations.

\begin{figure*}[t]
    \centering
    \includegraphics[width=1\linewidth]{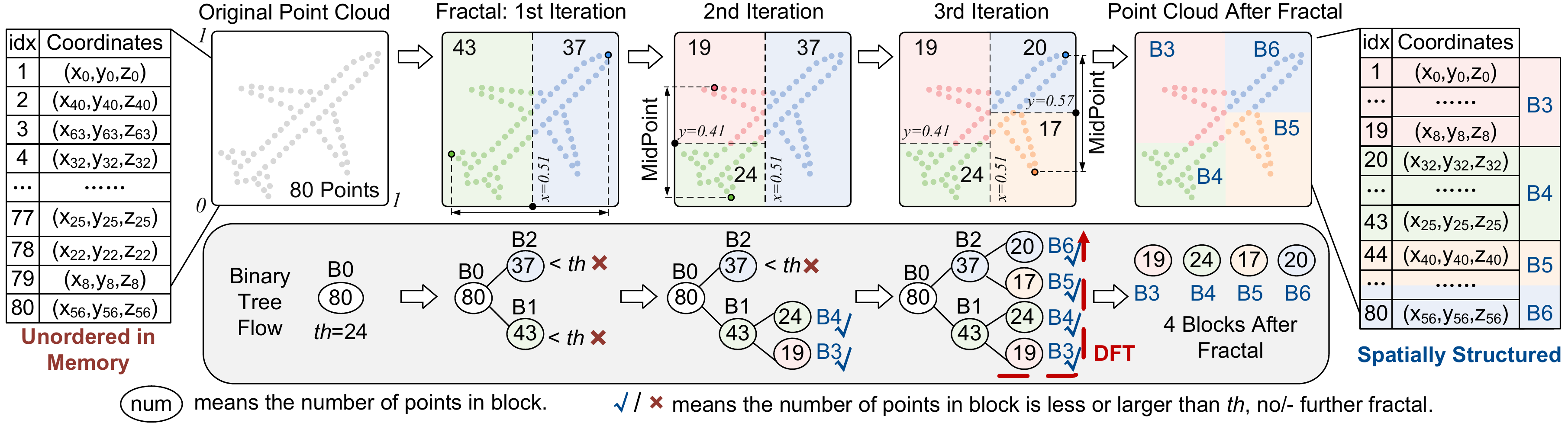}
    \caption{The workflow of the proposed Fractal method, where \textit{th} is the maximum points per block.}
    \label{fig:Fractal_workflow}
\end{figure*}

\textbf{Takeaway.}
The hardware efficiency of Fractal stems from its elimination of sorting overhead and adoption of fully decomposable traversal logic. With an inclusive, hardware-friendly traverser, \textit{\T} unlocks greater parallelism and improved scalability, making it well-suited for modern PNN accelerators targeting real-world, large-scale inputs.

Finally, \textit{\T} embodies a principled algorithm-hardware co-design philosophy. 
At the algorithmic level, it exploits object shape to achieve balanced partitioning with minimal complexity. 
At the architectural level, its traversal mechanism maps naturally to highly parallel hardware structures. 
Moreover, it establishes a foundation for decomposing and parallelizing all point operations, an aspect overlooked by prior accelerators, as further discussed in the next section.
This synergy between algorithmic insight and architectural design allows \textit{\T} to deliver high efficiency and scalability for large-scale PNN workloads.

\section{System Design: Fractal and Parallelization}

This section explains the system-level optimizations of \textit{\T}, which contains two techniques: the proposed Fractal method and the Block-Parallel Point Operations.

\subsection{Fractal: A Shape-Aware Partition}

\revision{RC1c}{Fractal is a hardware-driven, shape-aware partitioning method inspired by fractal geometry, featuring a DFT-based memory layout and threshold-controlled block division, which jointly balance hardware efficiency and network accuracy}. The core idea of Fractal is that point distributions in point clouds typically align with object shapes. Based on this, Fractal leverages shape information to adaptively partition point clouds into balanced blocks: dense regions are recursively partitioned into finer blocks, while sparse regions require fewer subdivisions. 
As shown in Fig.~\ref{fig:Fractal_workflow}, Fractal adopts a binary tree structure for both partitioning and storage. 
Given a point cloud and a predefined threshold $th$ (maximum points per block), Fractal iteratively partitions the input into two sub-blocks per step, expanding the binary tree through a recursive process, \revision{RA10a}{as illustrated in Alg.~\ref{alg:fractal} (Rows~4–7). 
Each iteration involves a single traversal to compute the midpoint from the current dimension’s extrema (Row~5)}, with dimensions cycled across iterations \revise{(Row~3)} to ensure balanced partitioning and low hardware complexity. 
Blocks exceeding $th$ are further partitioned, with smaller $th$ producing deeper, finer-grained partitions. 
Fractal organizes blocks in memory using depth-first traversal (DFT) \revise{(Row~8)}, \revision{RA2}{while points within each block remain randomly arranged without degrading performance. 
Before Fractal, the globally random layout of point clouds can cause multiple compute units to access different addresses within the same memory bank, leading to severe bank conflicts~\cite{feng2022crescent}. 
After Fractal, the point cloud is organized in blocks that are distributed across different banks.

Each block is exclusively assigned to a single compute unit with a dedicated memory interface, rendering the intra-block randomness irrelevant to performance
.}
Moreover, the DFT-based organization in Fractal ensures that adjacent memory blocks correspond to spatially adjacent regions, maintaining spatial continuity. 
This mitigates the control complexity of identifying neighboring blocks in conventional strategies~\cite{kim2021pnnpu,feng2022crescent}, and enables efficient memory access, as accessing neighboring regions simply involves reading consecutive memory blocks, facilitating further hardware optimizations.

\begin{algorithm}[t]
\caption{Fractal Partitioning}
\small
\begin{algorithmic}[1]
\Require \revision{RA10a}{Point cloud $P$, threshold $th$, dimension index $d$}
\Function{Fractal}{$P, d$}
    \If{$|P| \le th$} \Return $P$
    \Else
        \State $dim \gets (d \bmod 3)$ \Comment{\revise{Cycle over x, y, z}}
        \State $mid \gets (\max(P[dim]) + \min(P[dim])) / 2$
        \State $(P_l, P_r) \gets \text{split}(P, dim, mid)$
        \State \Return Node(\Call{Fractal}{$P_l, d{+}1$}, \Call{Fractal}{$P_r, d{+}1$})
    \EndIf
\EndFunction
\State $Blocks \gets \text{DFT\_traverse}(\Call{Fractal}{P, d=0})$
\end{algorithmic}
\label{alg:fractal}
\end{algorithm}

Fig.~\ref{fig:Fractal_workflow} illustrates an example of Fractal with an 80-point input point cloud and a threshold $th = 24$. In the first iteration, Fractal computes the midpoint of the x-dimension of input point cloud $B0$, and partitions the input into two blocks: $B1$ and $B2$, with (43, 37) points, respectively. Since the number of points in both blocks exceeds $th$, they both undergo further fractal. In the second iteration, $B1$ is split along the y-dimension into $B3$ (19 points) and $B4$ (24 points), both meeting the threshold and requiring no further partitioning. The process continues until all blocks contain fewer than $th$ points, forming a binary tree structure with four leaf nodes, representing the four point blocks after Fractal process. By leveraging shape information to adaptively partition with traversing, Fractal produces balanced blocks with low computational complexity. Its binary tree structure, organized in depth-first memory order, enables streamed memory access. Therefore, Fractal is hardware-friendly and scalable, making it efficient for large-scale point cloud processing.

\subsection{Block-Parallel Point Operations (BPPO)}

\begin{figure}[t]
    \centering
    \includegraphics[width=0.9\linewidth]{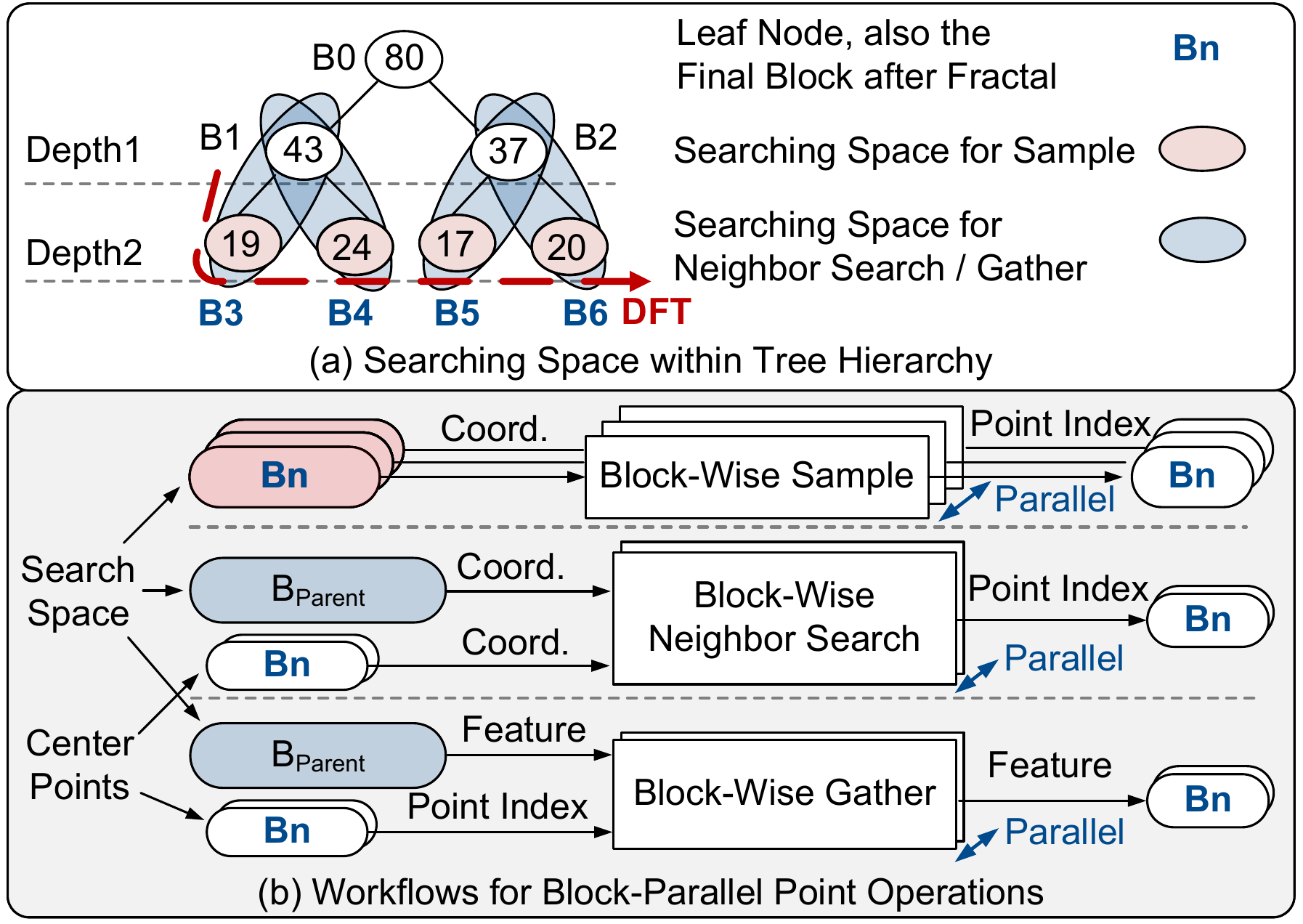}
    \caption{The (a) search space in tree hierarchy and (b) workflows of block-parallel point operations, including block-wise sampling, neighbor searching, and gathering.}
    \label{fig:SearchingSpace}
\end{figure}

After applying Fractal, the point cloud is adaptively partitioned into multiple balanced blocks and stored accordingly. Leveraging this strategy, we propose block-parallel point operations, which decompose and parallelize all sampling, neighbor searching, and gathering in PNNs by reducing their search space from entire point cloud to local blocks, as illustrated in Fig.~\ref{fig:SearchingSpace} with the 80-point point cloud in Fig.~\ref{fig:Fractal_workflow} as an example. 

\textbf{Block-Wise Sampling.} During sampling, conventional Farthest Point Sampling (FPS) iteratively searches the entire point cloud, incurring high computational overhead. We propose a block-wise sampling method that leverages the tree-structured point cloud storage from Fractal to decompose FPS. In block-wise sampling, instead of performing global searches, FPS is executed independently within each block, limiting the search space and computation to the current node in the tree structure, as shown in Fig.~\ref{fig:SearchingSpace}. Since leaf nodes are mutually independent, these block-wise operations can run in parallel, enabling inter-block parallelism and simultaneous sampling across multiple regions. The final sampled set is obtained by aggregating the results from all blocks.

A key challenge for applying FPS within blocks while maintaining accuracy is determining the appropriate number of sampled points per block, given the non-uniform point distribution. Prior works~\cite{kim2021pnnpu, zhou2023energy, zhou2024adjustable} rely on multiple hyperparameters and additional prediction steps to estimate sampling counts. In contrast, our block-wise sampling uses a fixed sampling rate across all blocks, avoiding such complexity without sacrificing accuracy. This is enabled by Fractal, which already aligns with point distribution and adaptively partitions point clouds into moderately balanced blocks, eliminating the need to track point distribution. Although block-wise sampling introduces slight deviations from global search, the impact on accuracy is negligible ($<0.2\%$) because Fractal preserves global geometry and neural networks tolerate minor variations. Consequently, block-wise sampling removes the global search in standard FPS while maintaining accuracy.

\textbf{Block-Wise Neighbor Searching.} Neighbor searching in PNNs consists of two key operations: grouping and interpolation, both require a search space and a center point cloud to identify neighboring points of the center point cloud within the search space. With Fractal, the point cloud is partitioned into blocks, enabling both block-wise grouping and block-wise interpolation, as illustrated in Fig. \ref{fig:SearchingSpace}. The center points correspond to leaf nodes in the point cloud tree. In this approach, if the leaf node is at depth 1, its search space is restricted to the same node. For deeper leaf nodes, \revision{RA4}{the search space expands to include their immediate parent nodes, which ensures a broader scope and is sufficient for maintaining network accuracy as shown in Fig.~\ref{fig:networkacc}.} 

Block-wise neighbor searching follows the same DFT order used in Fractal’s block storage, enabling data reuse and minimizing memory access. As illustrated in Fig.~\ref{fig:SearchingSpace}(a), the DFT starts from $B0$, first processing the leaf node $B3$ (depth 2). Its parent node $B1$ (depth 1, containing $B3$ and $B4$) is loaded on-chip to expand $B3$’s searching space. When moving to $B4$ (depth 2), its parent node $B1$ (depth 1) is already cached, allowing direct data reuse and eliminating redundant memory access. The same process applies to $B5$ and $B6$, where parent data are shared across sibling nodes without additional memory overhead. This DFT-based traversal supports a fully streamed block-wise neighbor searching pattern while avoiding additional parent-node access overhead. Moreover, this block-wise strategy relaxes the global-search dependencies of conventional neighbor searching, requiring only relevant blocks to be loaded on-chip. Consequently, all operations can be executed entirely on-chip, even for large-scale point cloud processing. 

\begin{figure}[tb]
    \centering
    \includegraphics[width=0.95\linewidth]{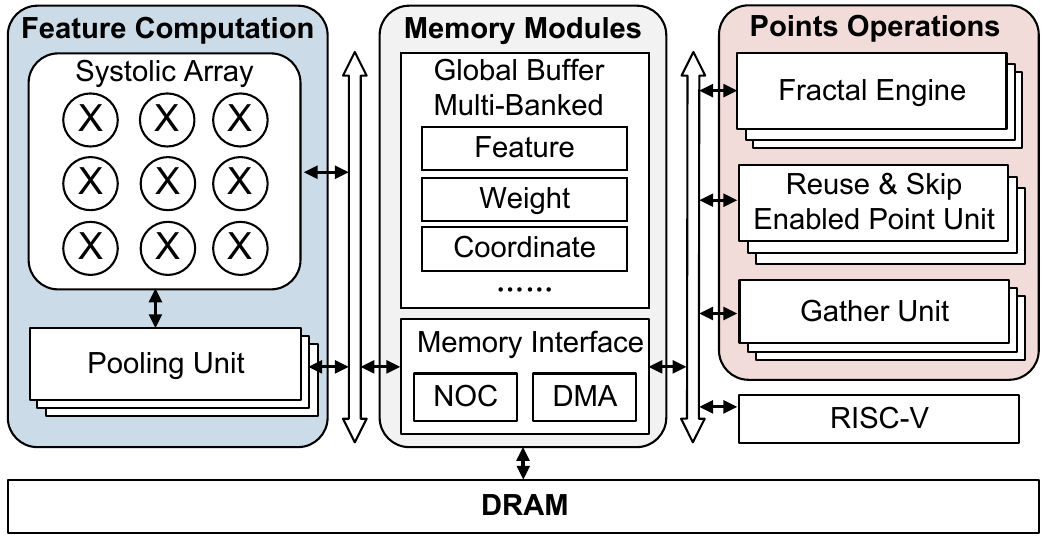}
    \caption{The overall architecture for \T.}
    \label{fig:overall architecture}
\end{figure}

\textbf{Block-Wise Gathering.}
Gathering retrieves feature values based on the point indices generated by neighbor searching. Conventional gathering requires a global search, meaning all feature data must be loaded on-chip. When the input size exceeds the on-chip buffer capacity, it incurs frequent DRAM accesses and substantial latency. In contrast, our block-wise gathering confines the search to relevant blocks, following the same search-space rule as block-wise neighbor searching. 
Enabled by Fractal, only related blocks need to be kept on-chip, enabling fully on-chip execution with limited buffer capacity and significantly reducing latency and energy consumption. \label{para:method-BWG}

\section{Hardware Architecture of FractalCloud}

This section presents the proposed hardware architecture of \textit{\T}, including the overall architecture design and detailed explanations of the submodules.

\subsection{Overall Architecture}\label{para:overall-archi}

\textit{\T} is a domain-specific accelerator designed to efficiently support PNNs across different input scales and diverse point operations. As illustrated in Fig.~\ref{fig:overall architecture}, the architecture comprises a fractal engine, reuse-and-skip-enabled point units (RSPUs), gathering units, pooling units, a systolic processing element (PE) array, a memory module, and a RISC-V core. The fractal engine implements all mainstream partitioning methods. RSPUs execute all point operations with block-parallel architecture supported. The gathering unit retrieves feature values based on the spatial indices, while the pooling unit performs max pooling operations in PNNs. The systolic PE array, configured as a $16\times16$ array following~\cite{feng2020mesorasi,lin2021pointacc,feng2022crescent}, handles convolutional and feedforward computations. The memory subsystem integrates a 274KB multi-bank global buffer following~\cite{lin2021pointacc} and a memory interface with a Network-on-Chip (NoC) for intra-chip communication and a Direct Memory Access (DMA) unit for DRAM interactions. A single-core six-stage RV32IMAC RISC-V processor, extended with Zifencei and additional control instructions, manages module configuration and data transfers.  \revision{RA6}{A lightweight configuration module between the RISC-V core and computation modules handles these instructions. The core writes control data into a buffer within this configuration module, which then segments and packages the data based on each computation module’s instruction length and dispatches it for execution.}

\textbf{Reconfigurability and Scalability.} Above hardware modules provide fundamental support for core operations in PNNs, enabling \textit{\T} to achieve high reconfigurability and scalability. Each module in \textit{\T} is equipped with dedicated instructions and configurable registers, allowing the RISC-V CPU to dynamically adjust its functionality and compose different point operations.
For example, the Fractal engine supports multiple partitioning strategies, including Fractal, uniform, and KD-tree-based methods, while the RSPUs and gather units can flexibly adapt to various sampling, grouping, and interpolation patterns. This modularity allows \textit{\T} to efficiently accommodate a wide range of PNN networks with different depths, resolutions, and sampling rates~\cite{deng2023pointvector, qian2022pointnext, qi2017pointnet++, qi2017pointnet,  phan2018dgcnn}.
Moreover, \textit{\T} scales efficiently across varying workloads. Large operations, such as large matrix multiplications, are automatically tiled by the compiler based on input size and hardware configurations. 
It generates a hardware-compatible storage allocation and configures the computing units, enabling seamless scaling with low overhead. 
Large-scale point operations are decomposed into fine-grained computations via the proposed Fractal method and executed in parallel following the BPPO, all scheduled and controlled directly on-chip. 
With such high reconfigurability and scalability, \textit{\T} achieves efficient acceleration for PNNs across different workloads, as shown in Fig.~\ref{fig:comparison}.

\begin{figure}[tb]
    \centering
    \includegraphics[width=0.9\linewidth]{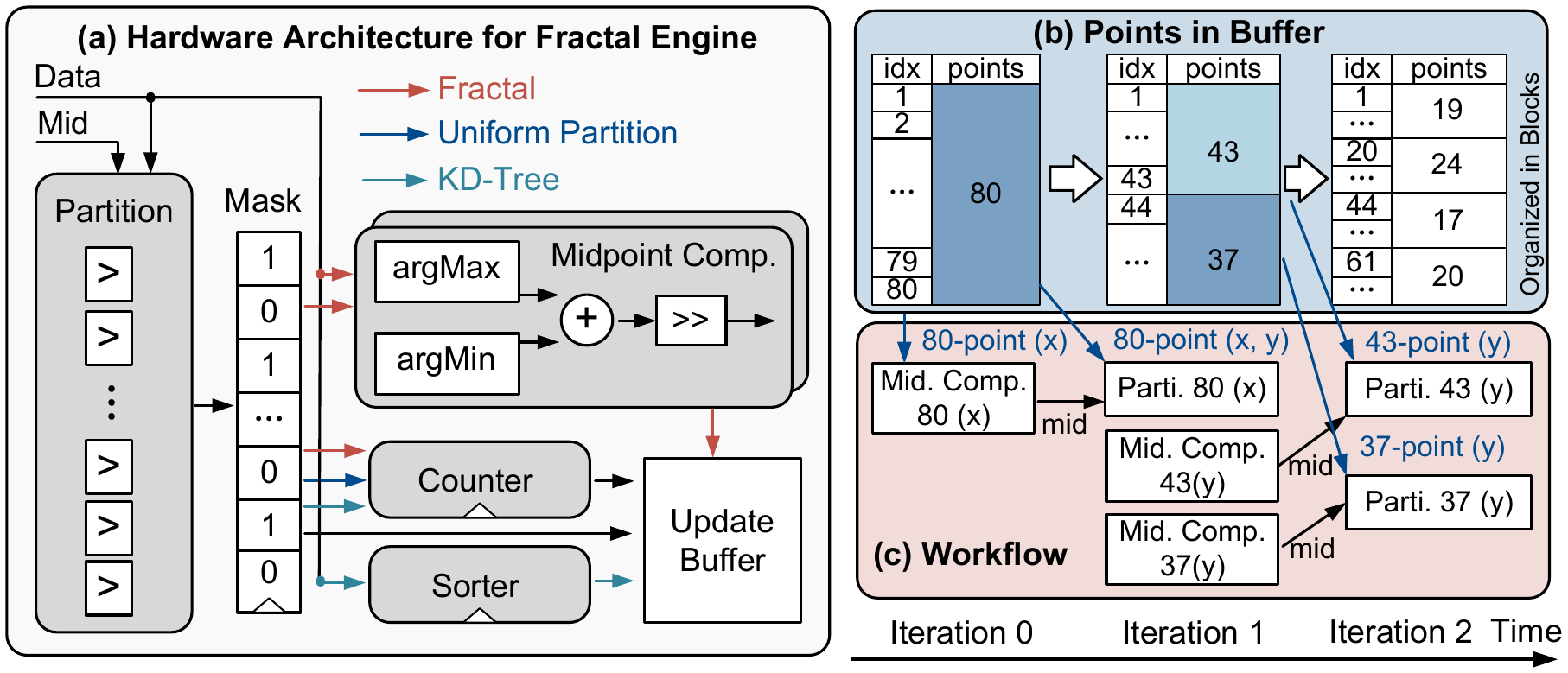}
    \caption{The (a) hardware architecture of fractal engine. (b) The optimized memory flow and (c) workflow for the Fractal.}
    \label{fig:PartitionArchi}
    \vspace{-0.5em}
\end{figure}

\subsection{Fractal Engine} 
The fractal engine efficiently supports the proposed Fractal method as well as mainstream partitioning methods such as uniform and KD-tree, as shown in Fig.~\ref{fig:PartitionArchi}(a). It comprises partition units, midpoint computation units, counters, and sorter units. Partition units use parallel comparators to partition points into sub-blocks based on the input Mid, while midpoint computation units calculate the midpoint of sub-block via min–max averaging with an addition and right-shift operation. Counters track point distribution after partitioning. And sorter units, based on a merge-sort hardware structure~\cite{lin2021pointacc}, provide the median for KD-tree partitioning. The dataflows of different partitioning methods are illustrated in Fig.~\ref{fig:PartitionArchi}(a), demonstrating the high hardware reuse of the fractal engine.

We have optimized the workflow for the proposed Fractal, as shown in Fig.~\ref{fig:PartitionArchi} (b) and (c) based on the same example of 80-point cloud with a threshold of 24 in Fig.~\ref{fig:Fractal_workflow}. The partition unit and midpoint computing unit operate in a pipelined manner to enhance throughput. In Iteration 0, the engine computes the midpoint along the x-dimension of the input point cloud. In Iteration 1, both x- and y-dimensions are processed: the x-dimension is used by the partition unit to divide points based on the previously computed midpoint, while the y-dimension is used by the midpoint computing unit to determine new midpoints for the resulting sub-blocks. After each iteration, the point cloud is partitioned into two sub-blocks based on mask values and stored in memory. Blocks with more points than the threshold (e.g., the 43-point and 37-point blocks) are further partitioned and passed to the next iteration. By Iteration 2, the entire point cloud is hierarchically partitioned, effectively optimizing the memory structure for subsequent processing.

\begin{figure}[t]
    \centering
    \includegraphics[width=0.9\linewidth]{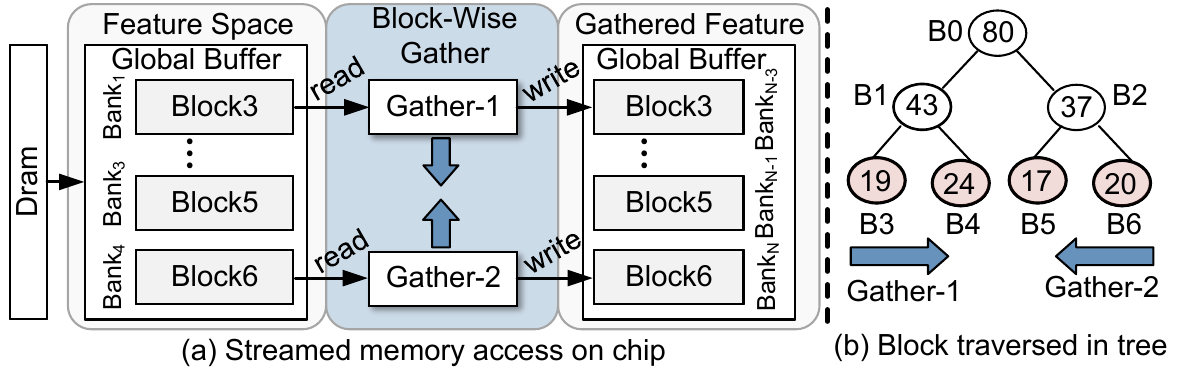}
    \caption{(a) The memory communication for block-wise gathering and (b) the traversing order of blocks in tree hierarchy.}
    \label{fig:gather}
\end{figure}

\textbf{Block-Wise Feature Access for Gathering.}
The proposed block-wise gather method narrows the search space from the entire point cloud to individual blocks, with each gather unit processing only its assigned block. Fig.~\ref{fig:gather} illustrates the memory access pattern for two gather units, where each block is stored in a separate memory bank. Gather Unit 1 performs a DFT traversal from left to right, processing Block 3 to Block 4, while Gather Unit 2 traverses in the opposite direction, processing Block 6 to Block 5. Both meet at the tree’s midpoint, completing the gathering process. For single-block gathering, the current block and its parent data are read from DRAM, and parent data are shared among all child nodes, increasing data reuse.

Block-wise gathering confines each unit to its designated memory bank, avoiding SRAM conflicts. Because only block-level data is retrieved, all necessary information fits in SRAM, eliminating DRAM lookups during processing. Moreover, thanks to organized data after Fractal, all DRAM accesses are also streamed rather than random, unlike conventional gathering. Detailed evaluations are provided in Section~\ref{para:ablation-BPPO}.

\subsection{Reuse-and-Skip-Enabled Point Unit (RSPU)}

\begin{figure}[t]
    \centering
    \includegraphics[width=0.9\linewidth]{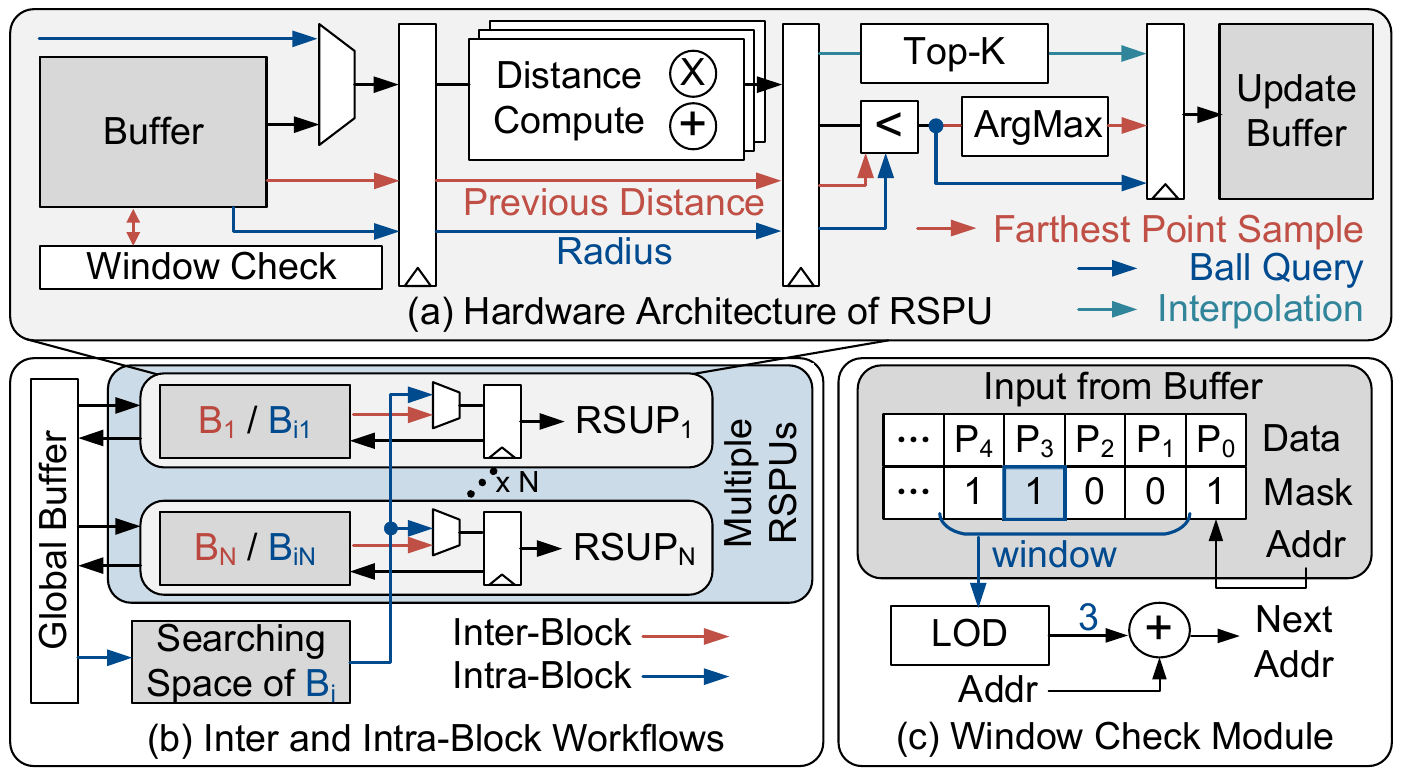}
    \caption{(a) The hardware design of reuse-and-skip-enabled point unit (RSPU) and (b) inter-block and intra-block workflow for multi-core RSPUs. (c) The hardware design for window check module, with a lowest-one detector (LOD) unit.}
    \vspace{-0.6em}
    \label{fig:RSPU}

\end{figure}

We propose multi-core reuse-and-skip-enabled point units (RSPU) that support both sampling and neighbor searching. It features a flexible block-level parallelism to enable efficient data reuse and redundant computation skipping, providing high hardware efficiency for block-parallel point operations. As shown in Fig.~\ref{fig:RSPU}(a), the RSPU comprises a windows check module, distance computing units, comparison units, an argmax unit, a top-k unit, and local buffers. RSPU operates in a pipeline manner.
The distance computing unit computes Euclidean distances. The comparison unit and argmax unit mainly incorporate comparators, while the argmax unit stores the maximum value from each comparison. The top-k unit employs an merge sort mechanism to compare and sort the input data against stored values, ultimately selecting the K smallest values from the input. 
The local buffer temporarily stores input/output data and mask information indicating whether each point has been sampled, ensuring efficient data management before writing to SRAM. The workflows of an RSPU for FPS, grouping, and interpolation are shown in Fig.~\ref{fig:RSPU}(a), demonstrating high hardware reuse.

\textbf{Flexible-Parallel Workflow of RSPUs.} Building on the RSPU design and BPPO, we propose a flexible-parallel workflow that exploits both intra-block and inter-block parallelism to accelerate point operations on multi-core RSPUs. 
\revision{RA10b}{The intra-block parallelism (blue lines in Fig.~\ref{fig:RSPU}(b), Row~6-8 in Alg.~\ref{alg:RSPU}) targets neighbor searching, with multiple RSPUs concurrently process different center points within a block while reusing search-space data from parent blocks, thereby saving memory accesses. 
The inter-block parallelism (red lines in Fig.~\ref{fig:RSPU}(b), Rows~2–3 in Alg.~\ref{alg:RSPU}) enables independent FPS execution on different blocks, achieving concurrent processing and avoiding redundant computations through a window-check mechanism discussed later}.

\textbf{Data Reuse of Coordinates.} As shown by the blue lines in Fig.~\ref{fig:RSPU}(b), intra-block workflows allow multiple RSPUs to perform neighbor searching concurrently with high data reusing. Specifically, each RSPU computes different center points within the same block while sharing the search space data from a dedicated buffer. This approach enhances data reuse by the number of RSPUs times, reducing redundant memory accesses for neighbor searching operations.

\begin{algorithm}[t]
\caption{{Flexible-Parallel Workflow on Multi-RSPUs}}
\label{alg:fractal_rspu}
\small
\begin{algorithmic}[1]
\Require \revise{Blocks $\{B_i\}$, parent block $B_p$, mode $m$, RSPUs $\{R_i\}$}

\If{$m=\text{FPS}$} \Comment{\revision{RA10b}{Inter-block parallelism}}
  \ForAll{$R_i$ \textbf{in parallel}}
    \State $B'_i \gets R_i(B_i, m)$
  \EndFor
\Else \Comment{\revise{Neighbor Search}}
  \ForAll{$B_i$}
    \State share $B_p$ data via buffer
    \ForAll{$R_i$ \textbf{in parallel}} \Comment{\revise{Intra-block parallelism}}
      \State $B'_{i,p_i} \gets R_i(p_i, B_p, m)$ \Comment{\revise{$p_i$: center points of $B_i$}}
    \EndFor
  \EndFor
\EndIf
\State \textbf{Update} $\{B'_i\}$ \textbf{in memory}
\end{algorithmic}
\label{alg:RSPU}
\end{algorithm}

\textbf{Computation Skipping Enabled by Window Check.} In standard FPS, each iteration performs a global traversal over all input points~\cite{lin2021pointacc}. However, already sampled points won't be sampled again, leading to redundant computations. To address this inefficiency, we introduce a window-check module that filters out sampled points (0s in Fig.\ref{fig:RSPU}(c)) before each iteration, ensuring only valid candidates (1s in Fig.\ref{fig:RSPU}(c)) participate in distance updates. As shown in Fig.\ref{fig:RSPU}(c), the module has a mask window input, indicating the sampling status of upcoming points. A lowest-one detector (LOD), implemented with a priority encoder, identifies the nearest ‘1’ to locate the next valid point and update the next memory address accordingly. This mechanism effectively bypasses redundant memory accesses and computations associated with already sampled points. Leveraging inter-block parallelism, where each FPS operates independently on its assigned RSPU, the window-check mechanism enables concurrent redundancy skipping across all parallel FPS operations.

\begin{table}[tb]
\centering
\small
\caption{Evaluated Networks and Datasets.}
\scalebox{0.85}{
\renewcommand{\arraystretch}{1.2} 
\begin{tabular}{lcccc}
\toprule[1.5pt]
\textbf{Model} & \textbf{Notation} & \textbf{Task} & \textbf{Dataset}  & \textbf{Scene}\\
\midrule[1.5pt]
{\makecell{PointNet++ \\[3pt] PointNeXt}} & {\makecell{PN++ (c) \\[3pt] PNXt (c)}} & Classification & ModelNet40 & Object\\
\midrule[0.5pt]
{\makecell{PointNet++ \\[3pt] PointNeXt}} & {\makecell{PN++ (ps) \\[3pt] PNXt (ps)}} & \makecell{Part Segmentation} & ShapeNet & Object\\
\midrule[0.5pt]
{\makecell{PointNet++ \\[3pt] PointNeXt \\[3pt] PointVector}} & {\makecell{PN++ (s) \\[3pt] PNXt (s) \\[3pt] PVr (s)}} & Segmentation & S3DIS & Indoor\\

\bottomrule[1.5pt]
\end{tabular}
}
\vspace{-0.5em}
\label{tab:evaluatedModels}
\end{table}

\section{Evaluation}

\subsection{Experimental Setup}
\label{exp_setup}

\textbf{Software Benchmarks.} We evaluate the proposed \textit{\T} using three advanced point cloud neural networks, PointNet++ \cite{qi2017pointnet++}, PointNeXt \cite{qian2022pointnext}, and PointVector~\cite{deng2023pointvector}, across three fundamental tasks: classification, part segmentation, and semantic segmentation, as summarized in Table \ref{tab:evaluatedModels}. To assess performance, we define two input scales: small-scale inputs (ranging from 1K to 4K points from ModelNet40 \cite{sun2022modelnet40}, ShapeNet \cite{chang2015shapenet}, and S3DIS \cite{armeni20163d} datasets), and large-scale inputs (ranging from 8K to 289K points from the S3DIS dataset). For accuracy benchmarking, we retrain and evaluate publicly available models from original repositories using PyTorch~\cite{imambi2021pytorch}, ensuring results comparable to those reported in the original papers. Classification performance is measured using overall accuracy (OA), whereas segmentation performance is evaluated using the mean Intersection-over-Union (mIoU) metric on the Area-5 test of the S3DIS dataset.

\textbf{Hardware Baselines.} To evaluate the proposed \textit{\T}, we employ two categories of hardware baselines. The first is a server-level GPU platform equipped with an \revision{RB1c}{}\revision{RC3}{NVIDIA TITAN RTX GPU, where all PNN implementations are CUDA-optimized using the Openpoints Library~\cite{openpoints}. The GPU is warmed up before measurement, and latency is obtained by averaging 20 synchronized inference runs measured with \texttt{time()}. Energy consumption of GPU is meatured using the NVIDIA Management Library~\cite{NVML}}. The second category includes state-of-the-art (SOTA) PNN accelerators: Mesorasi \cite{feng2020mesorasi}, PointAcc \cite{lin2021pointacc}, and Crescent \cite{feng2022crescent}. For a fair comparison, since Mesorasi and Crescent utilize the delayed-aggregation technique to reduce the latency of MLPs, we also integrate it into \textit{\T}. Moreover, as Mesorasi and Crescent do not support the FPS method in the sampling step, which is critical for large-scale point cloud processing, we equip them with the FPS engine from PointAcc. Key hardware specifications (e.g., frequency, cores, and data bit-width) are aligned with all SOTA accelerators to ensure fair comparisons, and all accelerators are evaluated under DDR4 memory for consistency.

\begin{table}[tb]
\centering
\small
\caption{Evaluated Hardware Accelerators.}
\scalebox{0.75}{
\renewcommand{\arraystretch}{1} 
\begin{tabular}{lcccc}
\toprule[1.5pt]
\makecell{\textbf{Accelerator}} & \textbf{Mesorasi}\cite{feng2020mesorasi} & \textbf{PointAcc}\cite{lin2021pointacc} & \textbf{Crescent}\cite{feng2022crescent} & \textbf{\textit{\T}}\\
\midrule[1.5pt]
\makecell{\textbf{Cores}} & 16x16 & 16x16 & 16x16 & 16x16 \\
\midrule[0.5pt]
\makecell{\textbf{SRAM (KB)}} & 1624 & 274 & 1622.8  & 274 \\
\midrule[0.5pt]
\makecell{\textbf{Frequency}} & 1GHz & 1GHz & 1GHz & 1GHz \\
\midrule[0.5pt]

\makecell{\textbf{Area (mm2)}} & 4.59 & 1.91 & 4.75 & 1.5 \\
\midrule[0.5pt]
\textbf{\makecell{DRAM \\ Bandwidth}} &\makecell{DDR4-2133\\17GB/s} & \makecell{DDR4-2133\\17GB/s} & \makecell{DDR4-2133\\17GB/s} & \makecell{DDR4-2133\\17GB/s} \\
\midrule[0.5pt]
\makecell{\textbf{Technology}} & 28nm & 28nm & 28nm & 28nm \\
\midrule[0.5pt]
\textbf{\makecell{Peak\\Performance}} & 512 GOPS & 512 GOPS & 512 GOPS & 512 GOPS \\
\bottomrule[1.5pt]
\end{tabular}
}
\label{tab:hardwareConfigs}
\end{table}

\begin{figure}[tb]
    \centering
    \includegraphics[width=1\linewidth]{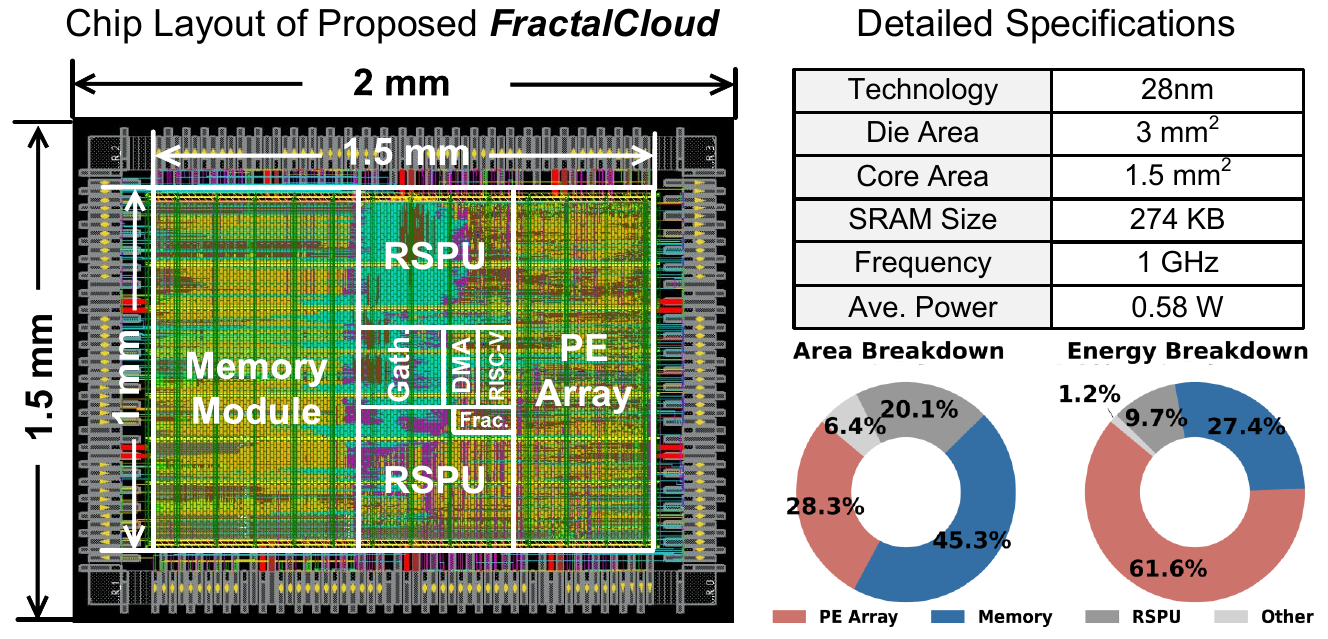}
    \caption{Chip layout, specifications, and breakdown of on-chip area and energy consumption for \textit{\T}.}
    \label{fig:post_layout}
\end{figure}

\textbf{Hardware Implementation.} The proposed \textit{\T} is implemented in Verilog HDL. We synthesize the hardware design using the Synopsys Design Compiler~\cite{design_compiler}, and place\&route using Cadence Innovus \cite{cadence_innovus} in \revise{TSMC} 28 nm technology. \revision{RA7}{Area for all modules are derived from post-layout reports, with SRAM generated by the memory compiler provided by TSMC. Power is measured through post-layout simulations using the real network data, ensuring more accurate estimation than prior synthesis-based and CACTI-based works~\cite{muralimanohar2009cacti, lin2021pointacc, feng2022crescent}.} We use 16-bit half-precision floating-point arithmetic for all computations to align with all SOTA works and preserve network accuracy. 
For end-to-end performance evaluation, we compare \textit{\T} with SOTA baselines using a cycle-level accurate simulator based on the open-source framework~\cite{guo2023olive}. The simulator is verified through RTL simulation. We do not simulate Mesorasi directly, as Crescent is an improved version of Mesorasi. Instead, the results for Mesorasi are sourced and scaled from the original paper. We use DRAMsim3 \cite{li2020dramsim3} to estimate the off-chip DRAM power consumption. To ensure a fair comparison, all designs are scaled to the 28nm technology node using the scaling method in \cite{fu2024softact}. The total core area of proposed \textit{\T} is $1.5 mm^2$, with an average power consumption of $0.58 W$, as shown in Fig.~\ref{fig:post_layout}. The hardware configurations of \textit{\T} and SOTA works are detailed in Table \ref{tab:hardwareConfigs}.

\begin{figure*}[t]
    \centering
    \subfloat{
        \includegraphics[width=0.95\linewidth]{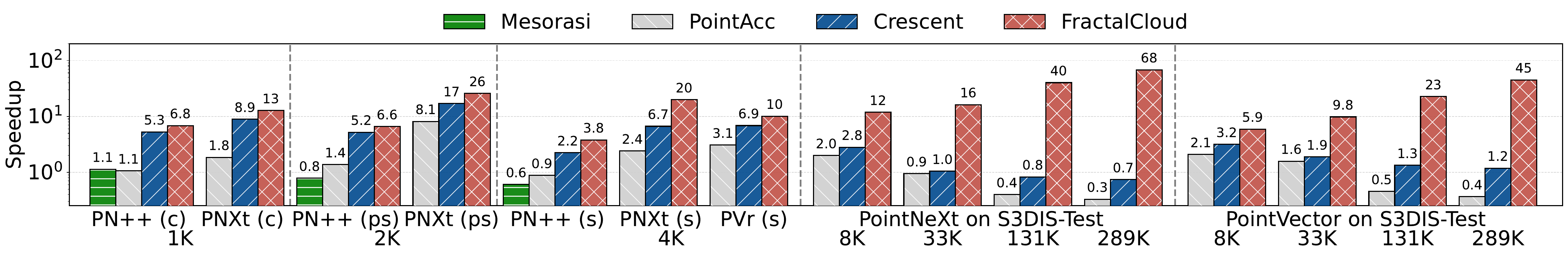}
    }\\
    \subfloat{
        \includegraphics[width=0.95\linewidth]{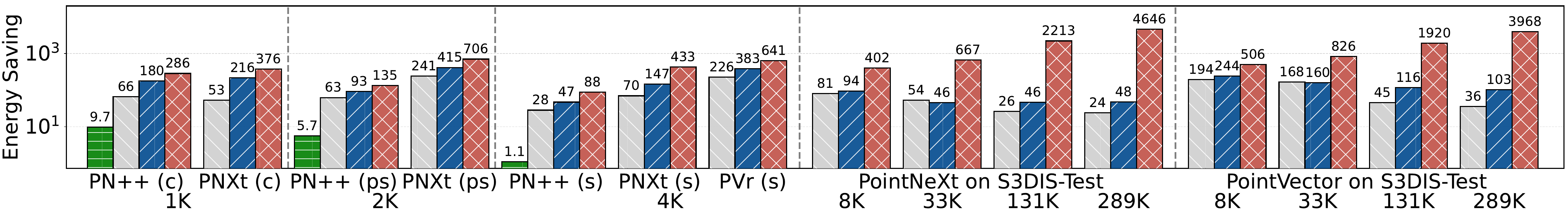}
    } 
    \caption{Comparison of speedup and energy saving achieved by Mesorasi, PointAcc, Crescent, and proposed \textit{\T}, normalized to GPU performance. Higher is better.}
    \label{fig:comparison}
\end{figure*}

\begin{figure}[tb]
    \centering
    \includegraphics[width=0.9\linewidth]{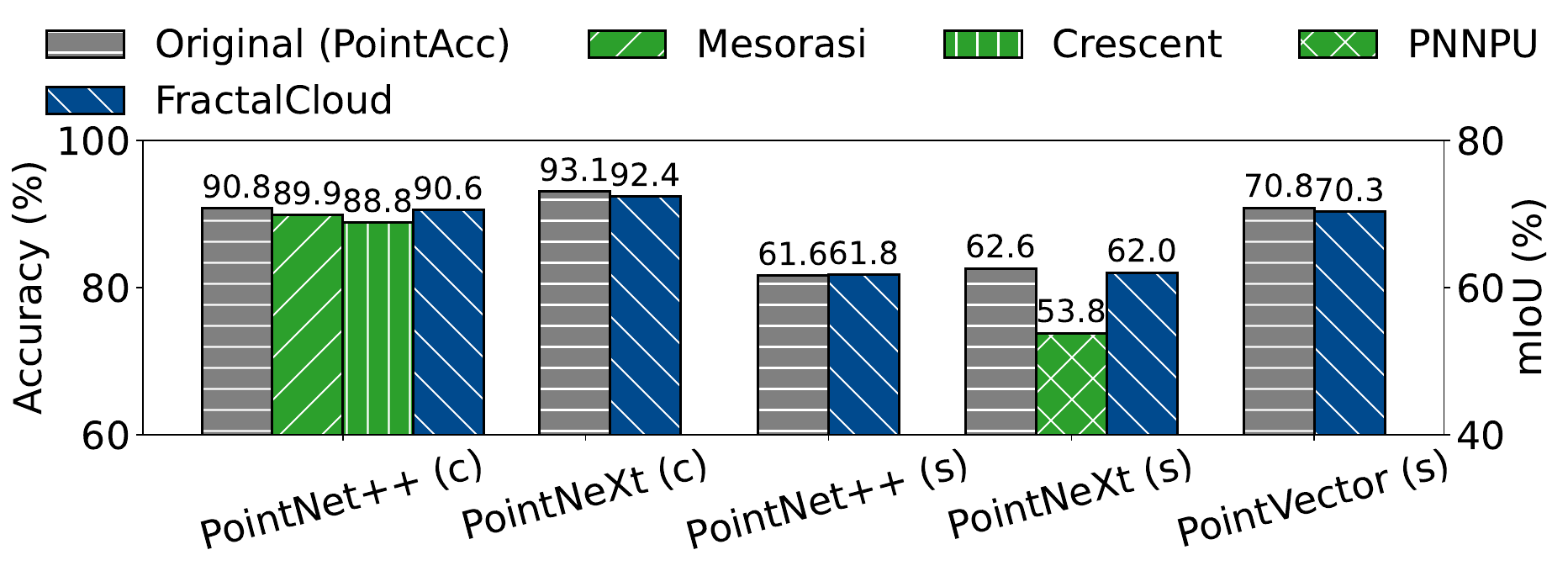}
    \caption{\recaption{RA8}{Network accuracy comparison} among: Original (PointAcc~\cite{lin2021pointacc}), Mesorasi~\cite{feng2020mesorasi}, Crescent~\cite{feng2022crescent}, PNNPU~\cite{kim2021pnnpu}, and \textit{\T} on different workloads. PointAcc is lossless.}
    \label{fig:networkacc}
\end{figure}

\subsection{Overall Results} \label{para:OverallResult}

\textbf{Network Accuracy.} We evaluate our proposed methods using PointNet++, PointNeXt, and PointVector models for classification and segmentation tasks, respectively, by replacing the original point operations with our proposed techniques and retraining networks, as shown in Fig. \ref{fig:networkacc}. The threshold for the proposed Fractal method is set to 64 for classification and 256 for segmentation. PointAcc~\cite{lin2021pointacc} preserves network accuracy but offers limited hardware improvement. Other prior works exhibit greater accuracy loss on the same tasks compared to \textit{\T}. Based on the same baseline, Mesorasi and Crescent have 0.9\%~\cite{feng2020mesorasi} and 2\%~\cite{feng2022crescent} loss for PointNet++ (c), respectively, due to network structure modifications and approximate memory access for improved hardware efficiency. PNNPU has 8.8\% loss for PointNeXt (s)~\cite{kim2021pnnpu, zhou2024adjustable} due to degraded feature representations after partitioning. In contrast, \textit{\T} achieves the best overall hardware performance, as shown in Fig.~\ref{fig:comparison}, while maintaining accuracy within 0.7\% loss across all models and input scales, demonstrating the best trade-off between hardware efficiency and model accuracy.

We further evaluate individual point operations to explain how \textit{\T} maintains high accuracy. Our findings indicate that directly applying the proposed block-wise sampling and block-wise interpolation preserves network accuracy, with less than 0.2\% accuracy loss. This is primarily because the blocks after Fractal align with the overall geometry of the point cloud, thereby maintaining the representational capacity of the points. 
Block-wise grouping introduces slight accuracy degradation, primarily due to numerical differences between local and original global searches. 
However, such errors can be mitigated through retraining \revision{RA8}{as evidenced by Fig.~\ref{fig:networkacc}}, since the extended search space in the block-wise grouping provides sufficient candidate points for neighbor selection. Block-wise gathering has no impact on network accuracy, as it does not modify original feature values. After retraining, all networks recover their accuracy, achieving performance almost identical to the original models.

\textbf{Speedup.} Fig.~\ref{fig:comparison} presents the inference speedup achieved by \textit{\T} compared to state-of-the-art accelerators, using GPU performance as the baseline. In the small-scale input scenario, \textit{\T} achieves speedups of 6.9×, 7.6×, 2.7×, and 19.4× over Mesorasi, PointAcc, Crescent, and GPU, respectively. For large-scale inputs, where point operations dominate latency, \textit{\T} demonstrates even more substantial improvements, with 63.4×, 27.8×, and 27.4× speedup over PointAcc, Crescent, and GPU. This performance gain primarily stems from three key optimizations. First, the Fractal method partitions the point cloud into multiple blocks with low hardware complexity, enabling balanced memory patterns for further optimization. 
Second, block-parallel point operations decompose and parallelize all point
operations from global search to local search. Third, the optimized hardware architecture enables fully parallel execution of both partitioning and point operations. Together, these optimizations ensure an average speedup of 21.7× compared to SOTA accelerators.\label{para:speedup}

\textbf{Energy Saving.} Fig.~\ref{fig:comparison} shows the energy savings of \textit{\T} under the same settings. In small-scale input scenarios, \textit{\T} achieves energy reductions of 44.3×, 4.1×, 1.9×, and 380× compared to Mesorasi, PointAcc, Crescent, and GPU, respectively. The improvements become more pronounced in large-scale inputs, with energy savings of 56.8×, 28.2×, and 1893× over PointAcc, Crescent, and GPU. Similar to the speedup results, these energy savings are rooted in \textit{\T}’s three optimizations. Mesorasi suffers from significant energy overhead from point operations, accounting for approximately 90\% of total energy consumption, due to the lack of optimization. PointAcc faces energy bottlenecks from excessive DRAM access caused by the global search problems and its limited 274KB on-chip buffer. Crescent mitigates DRAM energy consumption with a larger buffer and streamed memory access, but at the cost of increased SRAM energy usage, leading to only marginal gains compared to PointAcc. As a result, \textit{\T} achieves an average energy saving of 27× compared to SOTA accelerators.

\begin{figure}[tb]
    \centering
    \subfloat[(a) Latency breakdown.\label{fig:sub1}]{
        \includegraphics[width=0.45\linewidth]{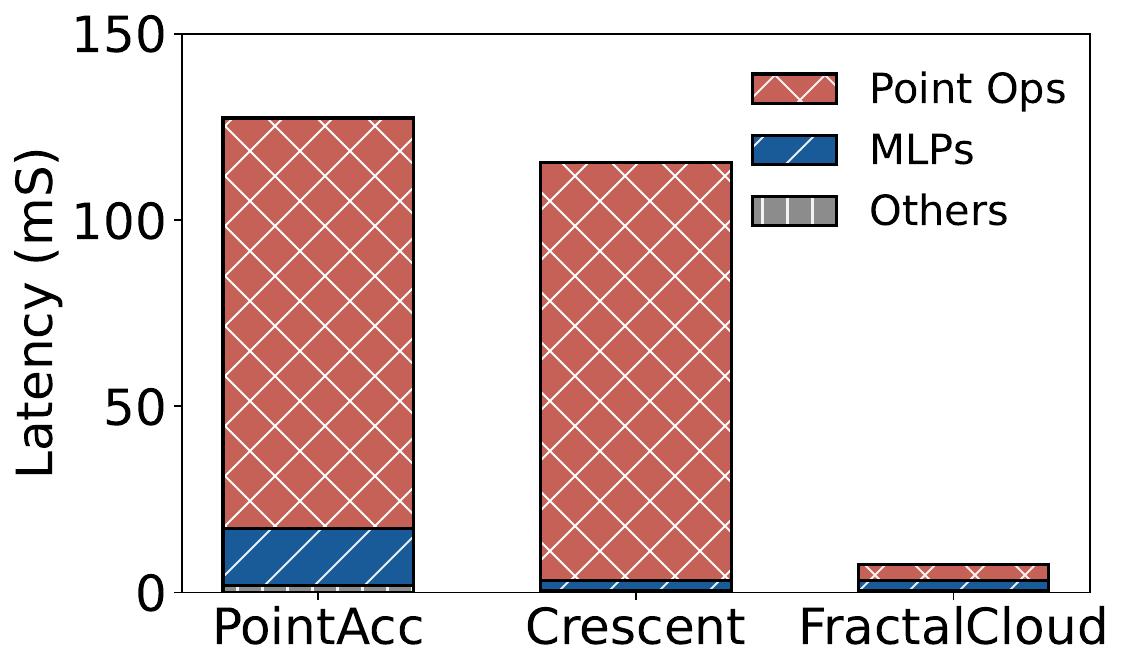}
    }
    \subfloat[(b) Energy breakdown.\label{fig:sub2}]{
        \includegraphics[width=0.45\linewidth]{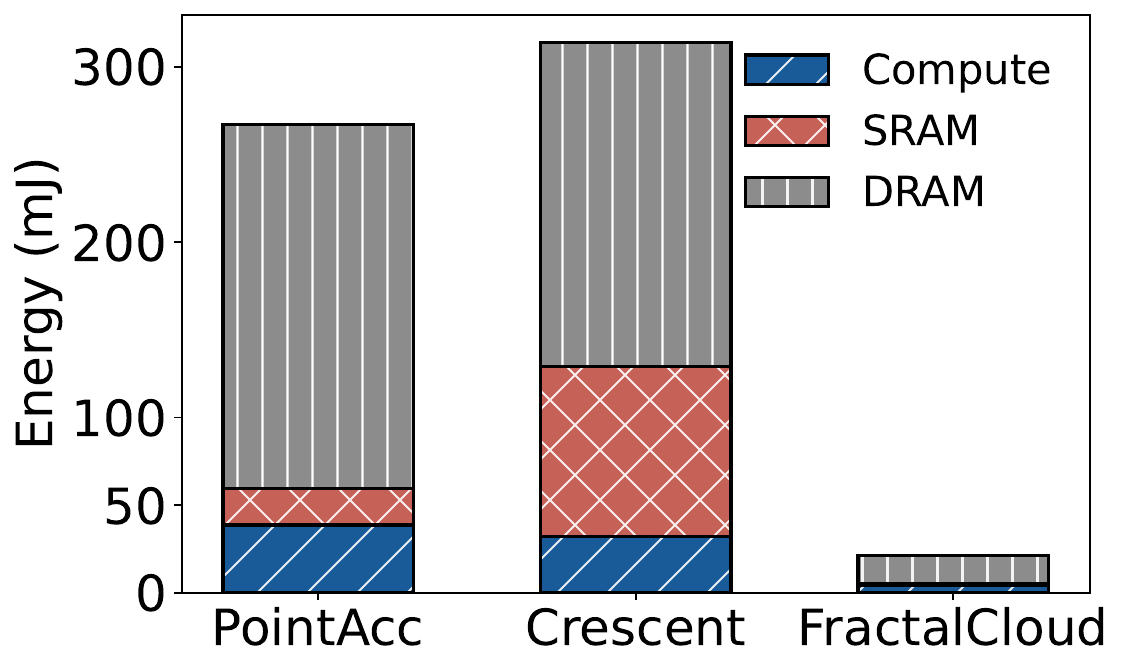}
    }
    \caption{(a) Latency breakdown and (b) energy breakdown for \textit{\T} and SOTA accelerators evaluated on PointNeXt segmenting S3DIS dataset.}
    \label{fig:latencyEnergyBreakdown}
    \vspace{-0.5em}
\end{figure}

\textbf{Latency and Energy Breakdown Analysis.}
Fig. \ref{fig:latencyEnergyBreakdown} presents a detailed breakdown of latency and energy consumption among PointAcc, Crescent, and \textit{\T} when executing PointNeXt on the S3DIS test dataset with 33K input points. In PointAcc, point operations rely on global search, requiring access and computation over the entire data. The small on-chip buffer cannot accommodate large-scale point cloud data, causing around 41\% of data to be retrieved from DRAM during point operations. These frequent off-chip accesses significantly increase both latency and energy consumption. Crescent reduces MLP latency by adopting the delay-aggregation method from~\cite{feng2020mesorasi}, but this increases the search space for gather operations, worsening the latency. To compensate, Crescent employs an improved KD-tree for memory streaming and a large 1622.8KB on-chip buffer. While this mitigates the problem of frequent DRAM access, it significantly increases SRAM power usage, and even causes a higher total energy consumption than PointAcc. As a result, Crescent achieves 1.1× speedup but 17\% more energy consumption compared to PointAcc. In contrast, \textit{\T} leverages Fractal to organize the point cloud, providing efficient partitioning for further optimization. BPPO reduces the search space of point operations to blocks, while RSPUs facilitate flexible and parallel hardware execution of these operations. Together, with 33K input points, these optimizations enable \textit{\T} to achieve an average of 16.2× latency reduction, 8.5× computation-related energy savings, and 14.7× memory-related energy savings, compared with PointAcc and Crescent, all within a limited 274KB buffer.

\subsection{Ablation Study on Proposed Techniques}

This section is organized as follows: We first compare the proposed Fractal with SOTA partitioning approaches and analyze the effect of different threshold settings. We then evaluate each point operation in BPPO to quantify each contribution. Finally, we assess the effectiveness of the proposed RSPU.

\begin{figure}[tb]
    \centering
    \includegraphics[width=0.9\linewidth]{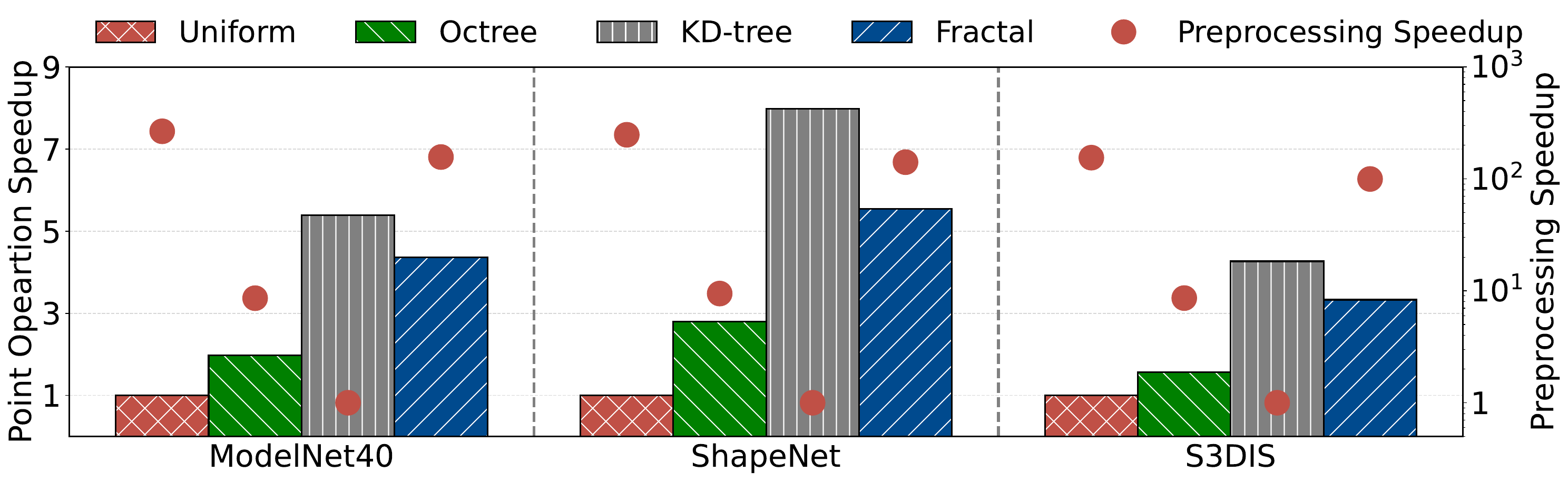}
    \caption{Speedup comparisons of point operation (bars, with uniform as baseline) and partitioning (dots, with KD-tree as baseline) among uniform, octree, KD-tree, and Fractal.}
    \label{fig:partitioncompare}
    \vspace{-0.5em}
\end{figure}

\textbf{Ablation Study on Fractal.} We compare Fractal with three representative partitioning techniques: uniform~\cite{kim2021pnnpu, zhou2023energy}, KD-tree~\cite{xu2019tigris,feng2022crescent}, and Octree partitionings~\cite{chen2023parallelnn, gao2024hgpcn, zhou202523}, as shown in Fig.~\ref{fig:partitioncompare}. Uniform partitioning offers minimal complexity but overlooks point cloud distribution, causing large variance within blocks and reducing point operation efficiency, with 8.8\% accuracy loss in large-scale tasks~\cite{zhou2024adjustable}. KD-tree achieves balanced blocks via recursive median selection but suffers from recursive and exclusive sorts, contributing up to 71\% of total latency when applied on \textit{\T}. Octree, as a uniform-based extension with dynamic subdivision, slightly reduces imbalance with increased control complexity~\cite{gao2024hgpcn}, but still incurs 3\% accuracy loss due to residual uniform partitioning~\cite{zhou202523, zhou2024adjustable}. Fractal addresses these limitations by adaptively partitioning points according to shape information, preserving geometric features while remaining hardware-friendly. It achieves 133× faster partitioning than KD-tree, 14.9× faster than Octree, and improves point operation performance by 2.1× over Octree and 4.4× over uniform partitioning. Additionally, Fractal incurs only 1\% area and 0.2\% energy overhead in end-to-end inference.

\begin{figure}[tb]
    \centering
    \includegraphics[width=0.75\linewidth]{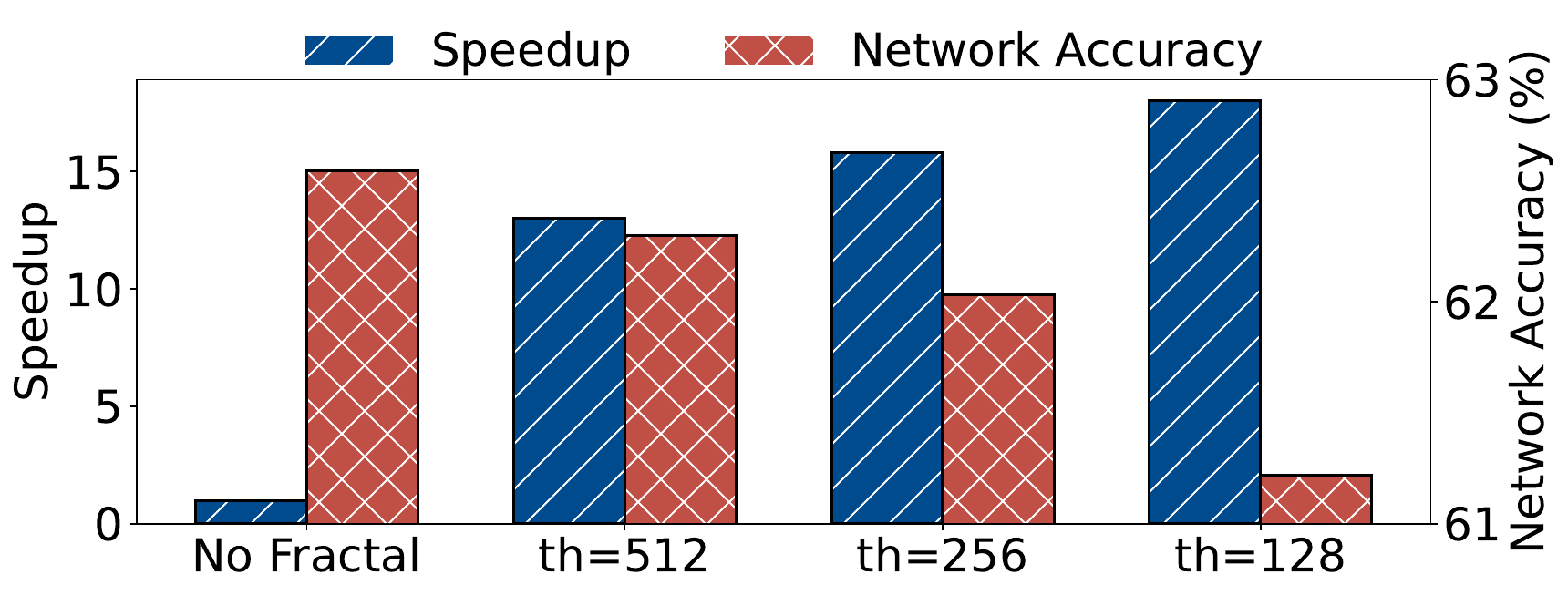}
    \caption{Impact of threshold selection in Fractal on hardware speedup and network accuracy for PointNeXt segmenting S3DIS dataset.}
    \label{fig:th_compare}
    \vspace{-0.5em}
\end{figure}

Threshold selection in Fractal trades off hardware speedup and accuracy, as shown in Fig.~\ref{fig:th_compare}. Overall, lower thresholds (\textit{th}) increase speedup by reducing points per block but may impair accuracy due to insufficient neighbors for feature extraction. 
\revision{CQ2}{Under extreme cases, excessive thresholds ($th{=}4k$) preserve accuracy but increase per-block computation complexity and memory traffic, with only 4.6$\times$ speedup. 
Overly small thresholds ($th{=}8$) over-partition the input, disrupting geometry, and causing random-like FPS and more errors in neighbor searching, causing over 8\% accuracy loss despite a 21$\times$ speedup. Through a greedy design-space exploration, \textit{th}=256/64 is chosen for large/small-scale inputs, achieving an optimal trade-off between speedup and accuracy}. 

\begin{figure}[t]
    \centering
    \includegraphics[width=0.95\linewidth]{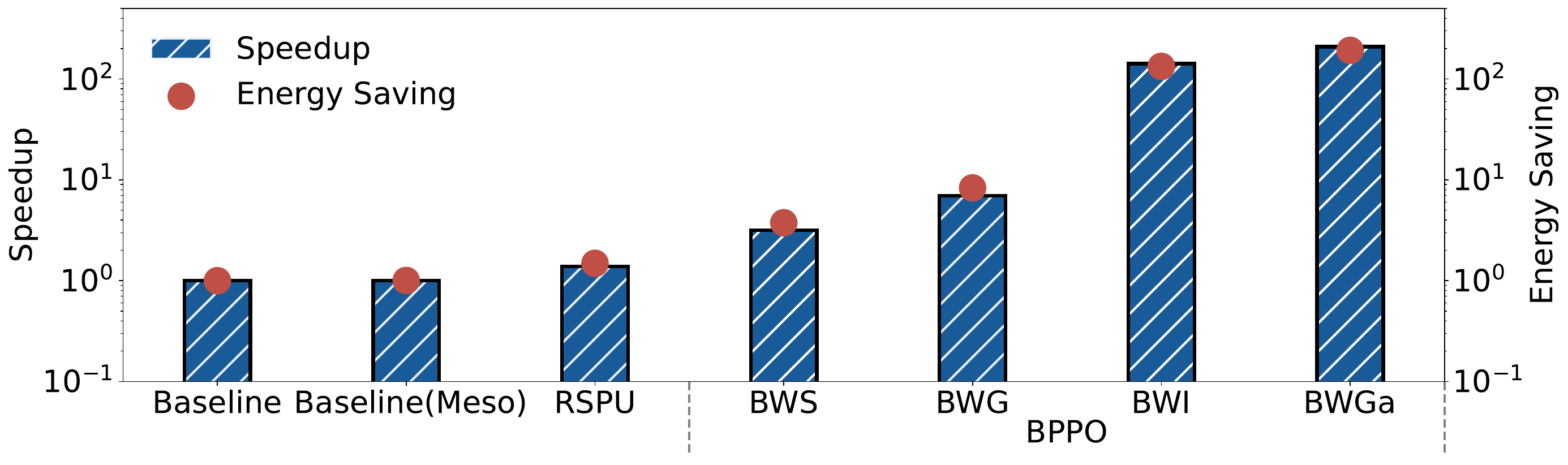}
    \caption{Comparison of speedup and energy savings after applying RSPU and each optimized point operation in BPPO.}
    \label{fig:performanceBreakdown}
    \vspace{-0.5em}
\end{figure}

\textbf{Ablation Study on Block-Parallel Points Operations (BPPO).}\label{para:ablation-BPPO} Fig.~\ref{fig:performanceBreakdown} presents the incremental speedup and energy savings of \textit{\T} during PointNeXt inference on the S3DIS dataset with 289K points. The \textit{Baseline} denotes \textit{\T} without optimizations, while \textit{Baseline (Meso)} applies Mesorasi’s delayed-aggregation\cite{feng2020mesorasi} for MLPs, offering only a marginal 1.004× gain. Adding RSPU eliminates redundant computations and improves data reuse in hardware, yielding 1.37× speedup and 1.48× energy savings over \textit{Baseline (Meso)}. BPPO further mitigates the global-search latency with parallel processing. Now sampling dominates 57\% of latency. Block-wise sampling (BWS) limits computation to local blocks, achieving 2.3× speedup and 2.5× energy savings. Grouping and interpolation then dominate latency (55\% and 43\%). Block-wise grouping (BWG) and interpolation (BWI) restrict searches to relevant blocks, providing 2.2× and 20× speedup and 2.2× and 16× energy savings, respectively. Finally, block-wise gathering (BWGa) alleviates DRAM fetches by enabling fully on-chip retrievals, adding 1.5× speedup and 1.4× energy savings. Overall, \textit{\T} achieves 209× speedup and 192× energy savings over both \textit{Baseline} and \textit{Baseline (Meso)}.

\textbf{Ablation Study on Reuse-and-Skip-Enabled Point Unit (RSPU).} \label{para:ablation-RSPU}
To efficiently accelerate point operations, we propose the reuse-and-skip-enabled point unit (RSPU), which supports block-parallel computing and integrates data reuse for neighbor searching and redundant computation skipping for sampling. Existing SOTA designs~\cite{lin2021pointacc, yoon2023efficient} mainly adopt point-level parallelism, where multiple input points are processed simultaneously to compute a single output point. Although this improves throughput, it requires complex hardware components (e.g., N-sorters, N-mergers) and cannot eliminate redundant computations in iterative FPS. Moreover, processing only one center point per iteration limits data reuse in neighbor searching, as search-space data cannot be effectively shared. In contrast, our RSPU employs a window-check mechanism to skip redundant FPS computations, achieving 3.6× speedup and 3.4× memory-access reduction over PointAcc. For neighbor searching, intra-block parallelism enables concurrent processing of multiple center points with the shared searching space, significantly improving data reuse with a 7.6× memory-access reduction. Overall, these optimizations improve PNN inference by 1.37× speedup and 1.48× energy efficiency.

\subsection{Discussion}

\revision{CQ1}{\textbf{Imbalance effect in Fractal.}
Outliers, irregular shapes, and different scenes in point cloud may cause partial imbalance in Fractal, but their impact on performance and memory load is minimal. 
Fractal enforces a threshold $th$ that limits the number of points per block. If an imbalance occurs, recursive partitioning continues until all blocks meet this constraint. 
Even under extreme shapes, such as two distant dense regions, the maximum imbalance among blocks after Fractal is strictly bounded by $th$, while spatial partitioning methods can reach the full input size in the worst case. 
Since both latency and memory load are dominated by the largest block, Fractal effectively constrains hardware imbalance. 
Moreover, Fractal cyclically utilizes all three spatial dimensions during partitioning, avoiding extreme cases in scene-level point clouds~\cite{armeni20163d} where coplanar points make one dimension non-splittable. 
Although outliers may lead to a few underfilled blocks, they represent only 0.5-2.5\% of points in S3DIS and are further mitigated by the threshold mechanism. 
Therefore, experiments on S3DIS show that, compared with a strictly balanced case, partial imbalance in Fractal increases overall latency by only 3.0\% and 2.8\% for PointNeXt and PointVector, respectively, confirming its negligible impact.}

\revision{RE2}{\textbf{Asymptotic Limitation.}
\textit{\T} can also accelerates extremely large-scale ($>500K$) point clouds, achieving an asymptotic speedup of 105.7× over GPU at 1M points on S3DIS with PointNeXt. But the accuracy may degrade if the block size is fixed, consistent with the trend shown in Fig.~\ref{fig:th_compare}, where larger block sizes can better preserve accuracy. This reflects a performance-accuracy tradeoff rather than a limitation of the Fractal itself. On the hardware side, the maximum supported input size mainly depends on DRAM capacity and the network configuration (e.g., a 24GB DRAM can handle 3M-point PointNeXt). In practice, however, existing PNNs rarely exceed 300K points, which already fully cover the needs of edge-level devices~\cite{roriz2024survey} targeted by our design.}

\revision{RE1}{\textbf{Potential Adaptations.} While \textit{\T} focuses on 3D point clouds, Fractal can also bring benefits to other data domains. In general, Fractal is applicable when (i) data preserve organized structures (e.g., graphs, tables), rather than being purely random, and (ii) local coherence exists among elements. For example, Fractal can exploit spatial locality in dynamic graphs to accelerate their construction and updates in DGCNN~\cite{phan2018dgcnn, wang2021object}, and may benefit high-dimensional tabular data~\cite{ukey2023survey} via locality-aware feature grouping and selection.}

\section{Conclusion}
The rapid growth of large-scale 3D point clouds presents new challenges and opportunities for efficient hardware acceleration of point cloud neural networks. To meet the emerging demands, we propose \textit{\T}, a fractal-inspired accelerator for efficient large-scale 3D point cloud processing. Specifically, \textit{\T} introduces the Fractal method to achieve balanced partitioning with low hardware complexity and maintained network accuracy. Block-parallel point operations are developed, decomposing and parallelizing all point operations from global search to local search. Moreover, a dedicated hardware design integrates on-chip fractal, flexible parallel processing, and optimized data reuse with redundancy elimination, maximizing the efficiency of our fractal approach. Implemented as a chip layout in 28nm technology, \textit{\T} achieves an average speedup of 21.7× and energy reduction of 27× compared to state-of-the-art PNN accelerators. 

\section*{Acknowledgments}
This work was supported in part by NSF grants 2328805 and 2112562, by ARO W911NF-23-2-0224, as well as the Beyond the Horizon Initiative through the Pratt School of Engineering, Duke University. The authors would like to thank the anonymous reviewers for their constructive feedback, which helped improve the quality of this work. We also extend our gratitude to Junyao Zhang, Haoxuan Shan, Mingyuan Ma, and Jonathan Ku for their technical support and insightful discussions.

\appendix
\section{Artifact Appendix}

\subsection{Abstract}

Our artifact contains the algorithmic implementation of \emph{FractalCloud}, enabling the evaluation of network performance under the proposed fractal-inspired point cloud processing architecture. The artifact includes the PyTorch implementation of the Fractal operation and Block-Parallel Point Operations, scripts to reproduce classification and segmentation results on ModelNet40 and S3DIS, and pretrained models for PointNet++, PointNeXt-S, and PointVector-L. The purpose of this artifact is to validate algorithmic correctness and network performance as reported in the paper.

\subsection{Artifact check-list (meta-information)}

{\small
\begin{itemize}
  \item {\bf Algorithm: } Fractal operation; block-parallel point operations; Point cloud neural networks
  \item {\bf Program: } Python (PyTorch implementation)
  \item {\bf Compilation: } N/A (Conda environment and Docker)
  \item {\bf Transformations: } N/A
  \item {\bf Binary: } Docker Image
  \item {\bf Model: } PointNet++, PointNeXt-S, PointVector-L
  \item {\bf Data set: } ModelNet40, S3DIS
  \item {\bf Run-time environment: } Ubuntu 24.04.3 LTS (Host), Docker container with Ubuntu 16.04.7 LTS, Python 3.7.16, PyTorch 1.10.1, CUDA 11.3
  \item {\bf Hardware: } x86-64 server with one NVIDIA TITAN RTX (24GB)
  \item {\bf Run-time state: } Standard inference with pretrained models
  \item {\bf Execution: } Python scripts
  \item {\bf Metrics: } Classification accuracy, segmentation mIoU
  \item {\bf Output: } Reported accuracy and segmentation metrics
  \item {\bf Experiments: } Classification on ModelNet40; segmentation on S3DIS
  \item {\bf How much disk space required (approximately)?: } $\sim$96GB total (including Docker, dependencies, datasets, pretrained models, and cache)
  \item {\bf How much time is needed to prepare workflow (approximately)?: } $\sim$20 minutes (Docker)
  \item {\bf How much time is needed to complete experiments (approximately)?: } 5--10 minutes for classification; 2--14 hours for segmentation
  \item {\bf Publicly available?: } Yes (\url{https://github.com/Yuzhe-Fu/FractalCloud})
  \item {\bf Code licenses (if publicly available)?: } MIT license
  \item {\bf Data licenses (if publicly available)?: } Original dataset licenses
  \item {\bf Workflow automation framework used?: } N/A
  \item {\bf Archived (provide DOI)?: } \url{https://zenodo.org/records/17851585}
\end{itemize}
}

\subsection{Description}

\subsubsection{How to access}

The source code is publicly available at: \url{https://github.com/Yuzhe-Fu/FractalCloud}.  
The artifact sources are also archived at Zenodo: \url{https://zenodo.org/records/17851585}.  
Pretrained weights for all evaluated models (PointNet++, PointNeXt-S, and PointVector-L) and inference logs for evaluated workflows are available at: \url{https://huggingface.co/YuzheFu/FractalCloud}.

\subsubsection{Hardware dependencies}

A server equipped with a GPU is required for evaluation. 
The artifact has been tested with an NVIDIA TITAN RTX GPU (24GB).

\subsubsection{Software dependencies}

The artifact relies on the following software components:
\begin{itemize}
    \item Host OS: Ubuntu 24.04.3 LTS
    \item Docker (optional): Docker 28.3.3
    \item Docker Image (provided): Ubuntu 16.04.7 LTS with all required dependencies preinstalled
    \item Python 3.7.16
    \item Pytorch 1.10.1
    \item CUDA 11.x (only required for local installation without the provided Docker image)
    \item Conda 4.9.2
\end{itemize}

\subsubsection{Data sets}

The evaluated datasets are:
\begin{itemize}
    \item ModelNet40 (classification)
    \item S3DIS (segmentation)
\end{itemize}

\subsubsection{Models}

\begin{itemize}
    \item PointNet++
    \item PointNeXt-S
    \item PointVector-L
\end{itemize}

Pretrained weights and inference logs (exact outputs reported in the paper) are available at \url{https://huggingface.co/YuzheFu/FractalCloud} for validating reproduced results.

\subsection{Installation}

Installation can be done through the provided Docker image or via running local setup scripts.  
A detailed step-by-step installation guide is provided in the project README file.

\subsection{Experiment workflow}

Example execution commands to reproduce classification and segmentation results are included in the README.  
Running the provided scripts will produce network accuracies matching the results reported in the paper.

\subsection{Evaluation and expected results}

The artifact is expected to reproduce the network accuracy metrics reported in the paper.  
Specifically, it validates the numerical consistency of the fractal implementation and its impact on network performance across multiple point cloud models.

\subsection{Experiment customization}

Users can modify experiment configurations, model architectures, and dataset paths by editing the script arguments and YAML configuration files as described in the project README.

\subsection{Notes}

N/A

\subsection{Methodology}

Submission, reviewing and badging methodology:

\begin{itemize}
  \item \url{https://www.acm.org/publications/policies/artifact-review-and-badging-current}
  \item \url{https://cTuning.org/ae}
\end{itemize}


\bibliographystyle{IEEEtran}
\bibliography{refs}

\end{document}